\documentclass[12pt]{article}
\usepackage{amssymb}
\usepackage{epsfig}
\def\vev#1{\left\langle #1\right\rangle}

\def\eq#1{{eq. (\ref{#1})}}
\def\vb#1{\vbox to #1 pt{}}
\def\21{$SU(2) \otimes U(1) $}
\def\np#1#2#3{           {Nucl. Phys. }{\bf #1} (19#2) #3}
\def\pl#1#2#3{           {Phys. Lett. }{\bf #1} (19#2) #3}
\def\pr#1#2#3{           {Phys. Rev. }{\bf #1} (19#2) #3}

\def\lsim{\raise0.3ex\hbox{$\;<$\kern-0.75em\raise-1.1ex\hbox{$\sim\;$}}}
\def\gsim{\raise0.3ex\hbox{$\;>$\kern-0.75em\raise-1.1ex\hbox{$\sim\;$}}}
\def\rp{$R_p \hspace{-1em}/\;\:$ }

\def\beqa{\begin{eqnarray}}
\def\eeqa{\end{eqnarray}}
\def\npb#1#2#3{{\it Nucl.\ Phys.\ }{\bf B #1} (#2) #3}

\def\hepph#1{{\tt hep-ph/#1}}

\newcommand {\app} [1] {Appendix~\ref{#1} }
\newcommand {\cosa} {\cos \alpha}
\newcommand {\sina} {\sin \alpha}
\newcommand {\cosaq} {\cos^2 \alpha}
\newcommand {\sinaq} {\sin^2 \alpha}

\newcommand {\chiz} [1] {\tilde{\chi}^{0}_{#1} }
\newcommand {\mchiz} [1] {m_{\tilde{\chi}^{0}_{#1} }}
\newcommand {\beq} {\begin{equation}}
\newcommand {\eeq} {\end{equation}}
\newcommand {\bea} {\begin{eqnarray}}
\newcommand {\eea} {\end{eqnarray}}
\newcommand{\mx}{\left[\begin{array}} 
\newcommand{\finmx}{\end{array}\right]} 
\newcommand {\fig} [1] {Fig.~\ref{#1}}
\newcommand {\figz} [2] {Fig.~(\ref{#1}{#2})}
\newcommand {\sect} [1] {Sec.~\ref{#1}}
\newcommand {\tab} [1] {Table~\ref{#1}}

\newcommand {\no} {\nonumber \\}

\begin{document}

\title{
\begin{flushright} \small \rm
  hep-ph/0011248  \\ 
  IFIC/00-62 
\end{flushright}
Testing neutrino mixing at future collider experiments}
\author{W.~Porod$^1$, M.~Hirsch$^{1,2}$, J.~Rom\~ao$^3$,
        and J.W.F.~Valle$^1$ \\[0.5cm]
 \small
$^1$Inst.~de F\'\i sica Corpuscular (IFIC), CSIC - U. de Val\`encia, \\ \small
Edificio de Institutos de Paterna, Apartado de Correos  22085\\ \small
E-46071--Val\`encia, Spain \\ \small
$^2$Department of Physics and Astronomy, 
                  University of Southampton,\\ \small
                  Highfield, Southampton SO17 1BJ,
                  England\\ \small
$^3$ Departamento de F\'\i sica, Instituto Superior T\'ecnico\\ \small
          Av. Rovisco Pais 1, $\:\:$ 1049-001 Lisboa, Portugal }
\maketitle

\begin{abstract}
  Low energy supersymmetry with bilinear breaking of R-parity leads to
  a weak-scale seesaw mechanism for the atmospheric neutrino scale and
  a radiative mechanism for the solar neutrino scale.  The model has
  striking implications for collider searches of supersymmetric
  particles. Assuming that the lightest SUSY particle is the lightest
  neutralino we demonstrate that (i) The neutralino decays inside the
  detector even for tiny neutrino masses.  (ii) Measurements of the
  neutrino mixing angles lead to {\it predictions} for the ratios of
  various neutralino decay branching ratios implying an independent
  test of neutrino physics at future colliders, such as the Large
  Hadron Collider or a Linear Collider.
  We study the lightest neutralino decay branching ratio predictions
  taking into account present supersymmetric particle mass limits as
  well as restrictions coming from neutrino physics, with emphasis on
  the solar and atmospheric neutrino anomalies.
\end{abstract}

\section{Introduction}
\label{sec:intro}

The simplest interpretation~\cite{latestglobalanalysis} of recent
solar and atmospheric neutrino data~\cite{solar,SuperK-atmos,atmos}
indicates that neutrinos are massive and, in contrast to the case of
quarks, at least one and possibly two of the lepton mixing angles are
large.  The possibility of testing for these angles at high energy
colliders seems very intriguing
\cite{acceltestold,nushort,acceltestnew} especially in view of the new
generation of colliders such as the LHC and a new high energy linear
collider.

One of the simplest ways to induce neutrino masses is if right-handed
neutrinos, singlets under the \21 SM gauge group exist, as expected in
a number of extended electroweak models and grand unified theories
(GUTs)~\cite{Schechter:1980gr,LR,threview}.  In this case there are
renormalizable neutrino Dirac masses similar to those of the charged
fermions and, in addition, potentially large Majorana mass terms for
the right-handed neutrinos. This leads to a neutrino mass matrix of
the form
\bea
\left( \begin{array}{cc} 0 & m \\ m & M \end{array} \right)
\eea
which has as eigenvalues
\bea
\lambda_{1,2} = \frac{M}{2} \mp \frac{\sqrt{M^2 + 4 m^2}}{2}
\eea
%
where $m = h^\nu \vev{H}$ with $h^\nu$ denoting a Dirac-type
Yukawa coupling for neutrinos and $\vev{H}$ the vacuum expectation
value (vev) of the SM Higgs field.  It is also assumed that right
handed neutrinos have a large mass term specified by $M$.
This so-called seesaw-mechanism \cite{seesaw} provides a general
recipe to generate neutrino masses.  For the simplest case $m/M \ll 1$
and just one generation one easily obtains that $\lambda_1 \simeq -
m^2/M$ while $\lambda_2 \simeq M + m^2/M$ corresponds to a heavy
right-handed state~\footnote{For detailed results on diagonalization
  of seesaw neutrino mass matrices see ~\cite{Schechter:1982cv}.}.
Typically the small neutrino masses required by the interpretation of
present neutrino data correspond to values of $M$ in a wide range
above $10^{9}$~GeV or so.
The parameters can be adjusted in such a way that they neatly explain
neutrino experiments but they are far from being predictive. They
mainly 'postdict' experimental data and lead to predictions, if any,
which are confined to the domain of ``neutrino'' experiments,
performed at underground installations, reactors or accelerators and
possibly neutrino factories. They hardly imply any signatures that may
be tested at high energy collider experiments such as the LHC.

An alternative possibility for providing neutrino masses exists in
which $M$ is of order $m_Z$ and that $m$ is thus rather small. It is
based on the idea of R-parity violation as origin of neutrino mass and
mixing~\cite{acceltestold,threview,rpold}. These models can have many
implications for gauge and Yukawa unification, low-energy physics
~\cite{epsrad,moreBRpV,beyond} as well as a variety of implications
for future collider experiments at high 
energies~\cite{acceltestnew,rpcoll,rpcoll2}.
The simplest model in which this is realized is an extension of the
Minimal Supersymmetric Standard Model (MSSM)~\cite{mssm} with bilinear
R-parity breaking terms in the superpotential~\cite{epsrad,moreBRpV}.

These bilinear terms lead to non-vanishing vevs for the sneutrinos
which in turn induce a mixing between neutrinos and
neutralinos~\cite{acceltestold,rpold}. This leads to an effective
neutrino mass matrix which is projective~\cite{acceltestold,Martin0nu}
implying that only one neutrino receives mass at the tree-level. This
provides for a way to account for the atmospheric neutrino problem.
For the explanation of the solar neutrino puzzle one has to perform a
1-loop calculation of the neutrino/neutralino mass matrix in order to
break the projectivity of the mass matrix~\cite{nushort,nulong}. The
net effect is a hybrid scheme in which the atmospheric scale is
generated by a weak-scale seesaw mechanism characterized by a mass
scale is of order $10^3$ GeV, while the solar neutrino scale arises
from genuine loop corrections~\cite{nushort,nulong}.

In the bilinear \rp model not only the neutrino masses but also the
neutrino mixing angles are predicted in terms of the three underlying
\rp parameters \cite{nushort,nulong}.  These same parameters also
determine the decay properties of the lightest supersymmetric particle
(LSP) which we assume is the lightest neutralino. This implies the
existence of simple correlations between neutrino mixing angles and
neutralino decay branching ratios, as already partly
observed~\cite{nushort,acceltestnew}. Note however that so far the
literature \cite{acceltestnew} has focused mainly on qualitative
statements based only on a tree-level approximation of the neutrino
mass and of the neutrino/neutralino couplings. First of all this is
not always reliable for extracting information on the atmospheric
angle. In addition, it is totally unsuitable for making any
determination of solar neutrino oscillation parameters from collider
physics.

In the present paper we present a quantitative approach to this
problem which takes into account the complete 1-loop calculation of
neutrino/neutralino masses and couplings. 
This is required to make reliable neutrino mass and mixing angle
predictions. Moreover such complete treatment is necessary in order to
establish reliable correlations between neutrino mixing angles and
neutralino decay branching ratios. This also includes the solar mixing
angle which has not yet been considered in this context in the
literature.
We discuss the most important correlations between neutrino physics
and neutralino physics in detail. In addition we give an overview of
all restrictions to the branching ratios of the lightes neutralino.
%
Moreover, we have taken into account also LSP
decays via real and virtual Higgs bosons, which have been so far
neglected in the literature.  Their contribution can be more important
than the one of the Z-boson if the lightest neutralino is mainly
gaugino-like as preferred by SUGRA scenarios and if third-generation
fermions are present in the final state.  This has some remarkable
implications as we are going to demonstrate.

Notice that we have ignored the results of the LSND
experiment~\cite{LSND} which would point to the existence of four
neutrinos in Nature~\cite{4nu} one of which is sterile. Actually it is
simple to extend our \rp model in order to include a light $SU(2)
\otimes U(1)$ singlet superfield~\cite{4nurp}. The fermion present in
this superfield is the sterile neutrino, which combines with one
linear combination of $\nu_e-\nu_{\mu}-\nu_{\tau}$ to form a Dirac
pair whose mass accounts for the LSND anomaly.  On the other hand the
sterile neutrino scalar partner can trigger the spontaneous violation
of R-parity, thereby inducing the necessary mass splittings to fit
also the solar and atmospheric neutrino data. This way the model can
explain all neutrino oscillation data leading to four predictions for
the neutrino oscillation parameters. However, for simplicity and for
definiteness we will focus here on the simpler scenario with only the
standard three light neutrinos.

The paper is organized as follows: In \sect{sec:model} we present the
model and discuss approximative formulas for some R-parity violation
couplings.  In \sect{sec:decays} we discuss the phenomenology of the
lightest neutralino at future accelerator experiments and in
\sect{sec:correlations} we discuss in detail the relationship between
neutrino-physics and neutralino-physics in this model. This includes
predictions for neutralino decays before SUSY is discovered and
various cross-checks after the SUSY spectrum is known. In
\sect{sec:conclusion} we draw our conclusions.  In
\app{app:approximate} we give some formulas for approximate
diagonalization of the generalized Higgs matrices.

\section{The model}
\label{sec:model}

R-parity conservation is an \texttt{ad hoc} assumption in the MSSM and
\rp may arise either as unification remnant~\cite{expl} or through \21
doublet left sneutrino vacuum expectation values (vevs)
$\vev{\widetilde{\nu_i}}$~\cite{rpold}.
Preferably we break R-parity spontaneously through \texttt{singlet
  right sneutrino vevs}, either by gauging L-number, in which case
there is an additional Z~\cite{ZR} or within the \21 scheme, in which
case there is a physical \texttt{majoron}~\cite{rpesp}.
In order to comply with LEP data on Z width we must assume that the
violation of R-parity is driven by \21 \texttt{singlet} sneutrino
vevs~\cite{SBRpV} since in this case the majoron has a suppressed
coupling to the $Z$.  Spontaneous R-parity violation may lead to
a successful electroweak baryogenesis~\cite{Multamaki:1998fu}.

As long as the breaking of R-parity is spontaneous then \texttt{only
  bilinear \rp terms arise in the effective theory} below the \rp
violation scale.  Bilinear R--parity violation may also be assumed
\texttt{ab initio} as the fundamental theory. For example, it may be
the only violation permitted by higher Abelian flavour
symmetries~\cite{Mira:2000gg}.  Moreover the bilinear model provides a
theoretically self-consistent scheme in the sense that trilinear \rp
implies, by renormalization group effects, that also bilinear \rp is
present, but \texttt{not} conversely.

The simplest \rp model (we call it \rp MSSM) is characterized by three
independent parameters in addition to those specifying the minimal
MSSM model.
Using the conventions of refs.~\cite{mssm,chargedhiggs} the model is
specified by the following superpotential \cite{epsrad},
\begin{equation}  
W=\varepsilon_{ab}\left[ 
 h_U^{ij}\widehat Q_i^a\widehat U_j\widehat H_u^b 
+h_D^{ij}\widehat Q_i^b\widehat D_j\widehat H_d^a 
+h_E^{ij}\widehat L_i^b\widehat R_j\widehat H_d^a 
-\mu\widehat H_d^a\widehat H_u^b 
+\epsilon_i\widehat L_i^a\widehat H_u^b\right] 
\label{eq:Wsuppot} 
\end{equation} 
where the couplings $h_U$, $h_D$ and $h_E$ are $3\times 3$ Yukawa
matrices and $\mu$ and $\epsilon_i$ are parameters with units of mass.
The second bilinear term in eq.~(\ref{eq:Wsuppot}) includes R--Parity
and lepton number violation in three generations.

Similary the soft supersymmetry breaking terms are obtained by adding
the corresponding R-parity breaking bilinear terms to the
supersymmetry breaking Lagrangian of the MSSM. For the explicit form
of these terms we refer to \cite{nulong}. The important point for the
following discussion is that beside the Higgs bosons also the
sneutrinos aquire non-zero vevs, which we denote be $v_D=<H^0_d>$,
$v_U=<H^0_u>$, $v_1=<\tilde \nu_e>$, $v_2=<\tilde \nu_\mu>$ and
$v_3=<\tilde \nu_\tau>$.
Note that the $W$ boson acquires a mass $m_W^2= g^2 v^2/4$, where
$v^2\equiv v_D^2 + v_U^2 + v_1^2+ v_2^2+ v_3^2 \simeq (246 \;
\rm{GeV})^2$. Like in the MSSM we define $\tan\beta=v_U/v_D$.

\subsection{Neutralino Mass matrix}

In the following we discuss the tree level structure of neutrino
masses and mixings 
as needed for the following discussion. 
A complete discussion of the 1-loop mass matrix
and the other mass matrices in this model can be found in
\cite{nulong}.  
%
In the basis $\psi^{0T}=
(-i\lambda',-i\lambda^3,\widetilde{H}_d^1,\widetilde{H}_u^2, \nu_{e},
\nu_{\mu}, \nu_{\tau} )$ the neutral fermion mass matrix ${\mathbf M}_N$
is given by
\beq
{\bold M}_N=\left[  
\begin{array}{cc}  
{\cal M}_{\chi^0}& m^T \cr
\vb{20}
m & 0 \cr
\end{array}
\right]
\eeq
where
\beq
{\cal M}_{\chi^0}\hskip -2pt=\hskip -4pt \left[ \hskip -7pt 
\begin{array}{cccc}  
M_1 & 0 & -\frac 12g^{\prime }v_D & \frac 12g^{\prime }v_U \cr
\vb{12}   
0 & M_2 & \frac 12gv_D & -\frac 12gv_U \cr
\vb{12}   
-\frac 12g^{\prime }v_D & \frac 12gv_D & 0 & -\mu  \cr
\vb{12}
\frac 12g^{\prime }v_U & -\frac 12gv_U & -\mu & 0  \cr
\end{array}  
\hskip -6pt
\right] 
\eeq
is the standard MSSM neutralino mass matrix and
\beq
m=\left[  
\begin{array}{cccc}  
-\frac 12g^{\prime }v_1 & \frac 12gv_1 & 0 & \epsilon _1 \cr
\vb{18}
-\frac 12g^{\prime }v_2 & \frac 12gv_2 & 0 & \epsilon _2  \cr
\vb{18}
-\frac 12g^{\prime }v_3 & \frac 12gv_3 & 0 & \epsilon _3  \cr  
\end{array}  
\right] 
\eeq
characterizes the breaking of R-parity.  The mass matrix ${\bold M}_N$
is diagonalized by 
\beq
{\cal  N}^*{\bold M}_N{\cal N}^{-1}={\rm diag}(m_{\chi^0_i},m_{\nu_j})
\label{chi0massmat}
\eeq
where $(i=1,\cdots,4)$ for the neutralinos, and $(j=1,\cdots,3)$ for
the neutrinos.  

We are interested in the case where the neutrino mass which is chosen
at tree level is small, since it will be determined in order to
account for the atmospheric neutrino anomaly. Under this assumption
one can perform a perturbative diagonalization of the
neutrino/neutralino mass matrix~\cite{Schechter:1982cv}, by
defining~\cite{Martin0nu}
\beq
\xi = m \cdot {\cal M}_{\chi^0}^{-1}
\label{defxi}
\eeq
If the elements of this matrix satisfy $\xi_{ij} \ll 1$ $\forall {ij}$
then one can use it as expansion parameter in order to find an
approximate solution for the mixing matrix ${\cal N}$.  Explicitly we
have
\begin{eqnarray}
\xi_{i1} &=& \frac{g' M_2 \mu}{2 det({\cal M}_{\chi^0})}\Lambda_i \cr
\vb{20}
\xi_{i2} &=& -\frac{g M_1 \mu}{2 det({\cal M}_{\chi^0})}\Lambda_i \cr
\vb{20}
\xi_{i3} &=& - \frac{\epsilon_i}{\mu} + 
          \frac{(g^2 M_1 + {g'}^2 M_2) v_U}
               {4 det({\cal M}_{\chi^0})}\Lambda_i \cr
\vb{20}
\xi_{i4} &=& - \frac{(g^2 M_1 + {g'}^2 M_2) v_D}
               {4 det({\cal M}_{\chi^0})}\Lambda_i
\label{xielementos}
\end{eqnarray}
where
\beq
\Lambda_i = \mu v_i + v_D \epsilon_i 
\label{lambdai}
\eeq
are the ``alignment'' parameters. From \eq{xielementos} and
\eq{lambdai} one can see that $\xi=0$ in the MSSM limit where
$\epsilon_i=0$ and $v_i=0$.  In leading order in $\xi$ the mixing
matrix ${\cal N}$ is given by,
\beq
{\cal N}^* \hskip -2pt =\hskip -2pt  \left(\hskip -1pt
\begin{array}{cc}
N^* & 0\\
0& V_\nu^T \end{array}
\hskip -1pt
\right)
\left(
\hskip -1pt
\begin{array}{cc}
1 -{1 \over 2} \xi^{\dagger} \xi& \xi^{\dagger} \\
-\xi &  1 -{1 \over 2} \xi \xi^\dagger
\end{array}
\hskip -1pt
\right) 
\eeq
%
%
%
The sub-matrices $N$ and $V_{\nu}$ diagonalize
${\cal M}_{\chi^0}$ and $m_{eff}$
\beq
N^{*}{\cal M}_{\chi^0} N^{\dagger} = {\rm diag}(m_{\chi^0_i}),
\eeq
\beq
V_{\nu}^T m_{eff} V_{\nu} = {\rm diag}(0,0,m_{\nu}),
\eeq
where 
\bea
m_{eff} &=& 
\frac{M_1 g^2 \!+\! M_2 {g'}^2}{4\, det({\cal M}_{\chi^0})} 
\left(\hskip -2mm \begin{array}{ccc}
\Lambda_e^2 
\hskip -1pt&\hskip -1pt
\Lambda_e \Lambda_\mu
\hskip -1pt&\hskip -1pt
\Lambda_e \Lambda_\tau \\
\Lambda_e \Lambda_\mu 
\hskip -1pt&\hskip -1pt
\Lambda_\mu^2
\hskip -1pt&\hskip -1pt
\Lambda_\mu \Lambda_\tau \\
\Lambda_e \Lambda_\tau 
\hskip -1pt&\hskip -1pt 
\Lambda_\mu \Lambda_\tau 
\hskip -1pt&\hskip -1pt
\Lambda_\tau^2
\end{array}\hskip -3mm \right) 
\eea
and
%
\bea
\label{mnutree}
m_{\nu} &=& Tr(m_{eff}) = 
\frac{M_1 g^2 + M_2 {g'}^2}{4\, det({\cal M}_{\chi^0})} 
|{\vec \Lambda}|^2.
\eea

Due to the projective nature of $m_{eff}$, only one neutrino acquires
mass~\cite{acceltestold}. As a result one can rotate away one of the
three angles in the matrix $V_{\nu,\mathrm{tree}}$, leading
to~\cite{Schechter:1980bn}
\beq
V_{\nu,\mathrm{tree}}= 
\left(\begin{array}{ccc}
  1 &                0 &               0 \\
  0 &  \cos\theta_{23} & -\sin\theta_{23} \\
  0 &  \sin\theta_{23} & \cos\theta_{23} 
\end{array}\right) \times 
\left(\begin{array}{ccc}
  \cos\theta_{13} & 0 & -\sin\theta_{13} \\
                0 & 1 &               0 \\
  \sin\theta_{13} & 0 & \cos\theta_{13} 
\end{array}\right) ,
\eeq
where the mixing angles can be expressed in terms of the {\it
  alignment vector} ${\vec \Lambda}$ as
\beq
\label{tetachooz}
\tan\theta_{13} = - \frac{\Lambda_e}
                   {(\Lambda_{\mu}^2+\Lambda_{\tau}^2)^{\frac{1}{2}}},
\eeq
\beq
\label{tetatm}
\tan\theta_{23} = - \frac{\Lambda_{\mu}}{\Lambda_{\tau}} \,.
\eeq
As discussed in detail in \cite{nulong} the inclusion of 1-loop
corrections to the mass matrix lifts the degeneracy between these
states. Only after including these corrections one obtains a
meaningful angle in the $1-2$ sector. Both features are required to
account for the solar neutrino data.

\subsection{Approximate Formulas for neutralino couplings}
\label{sec:approx}
\begin{figure}[t]
 \setlength{\unitlength}{1mm}
\begin{center}
\begin{picture}(140,80)
\put(-35,-112){\mbox{\epsfig{figure=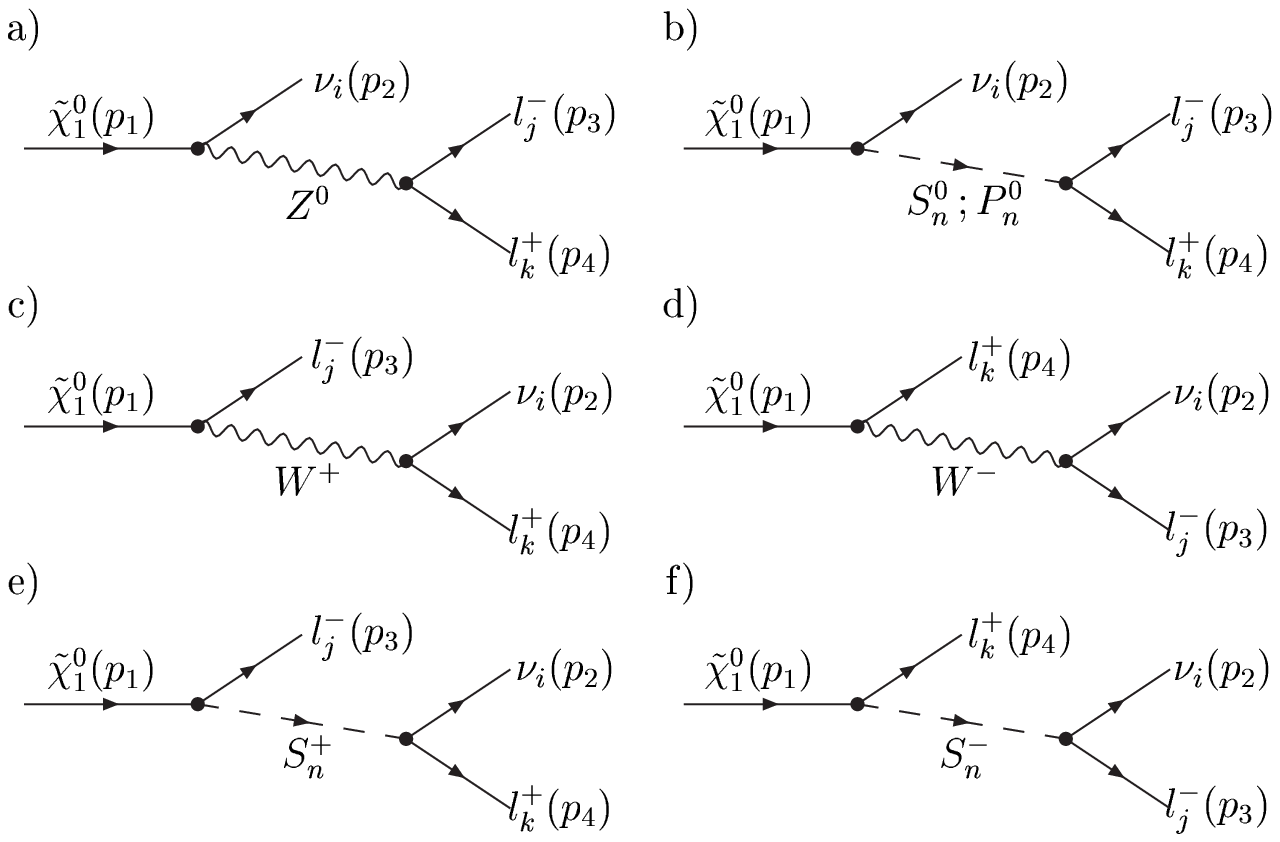,height=29.7cm,width=21.cm}}}
\end{picture}
\end{center}
\caption[]{ Feynman graphs for the decay 
  $\chiz{1} \to \nu_i \, l^-_j \, l^+_k$.}
\label{fig:graphs}
\end{figure}

\begin{figure}[ht] 
 \setlength{\unitlength}{1mm}
\begin{center}
\begin{picture}(140,80)
\put(-35,-112){\mbox{\epsfig{
               figure=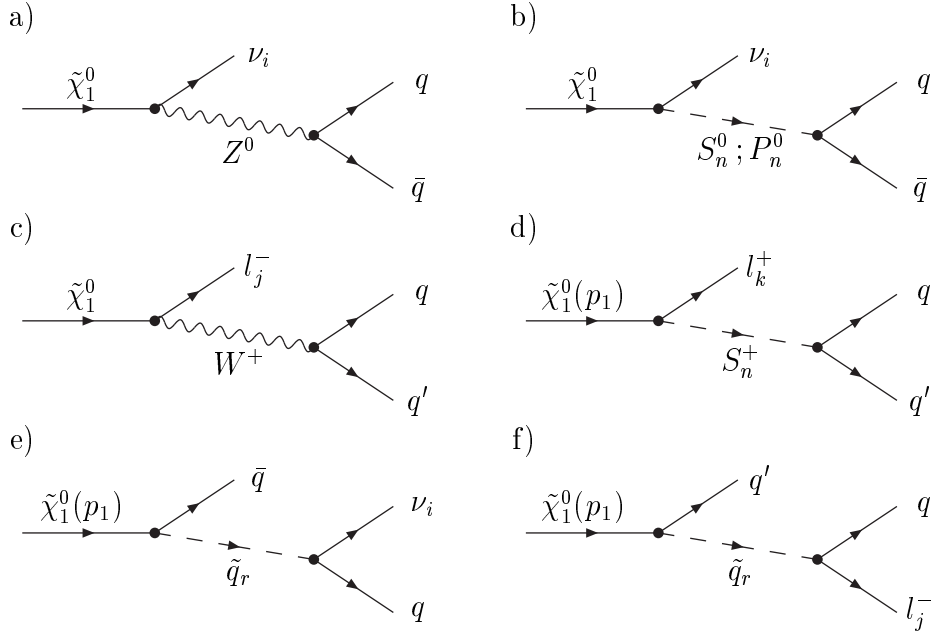,height=29.7cm,width=21.cm}}}
\end{picture}
\end{center}
\caption[]{ Generic Feynman graphs for semi-leptonic neutralino decays.}
\label{fig:graphs1}
\end{figure}
\begin{figure}[ht] 
 \setlength{\unitlength}{1mm}
\begin{center}
\begin{picture}(140,30)
\put(-35,-140){\mbox{\epsfig{
               figure=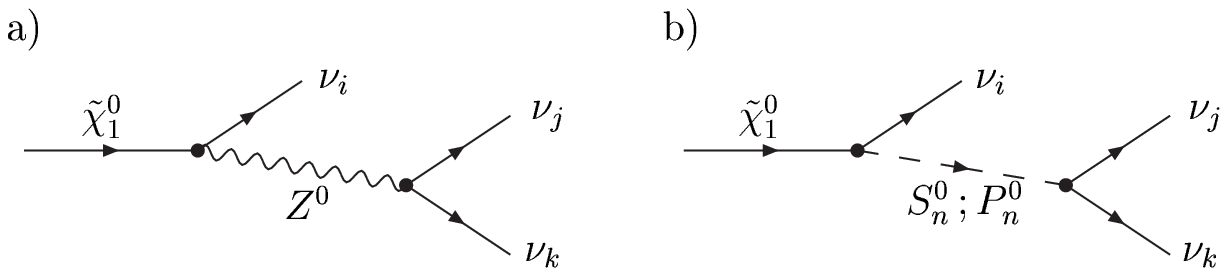,height=29.7cm,width=21.cm}}}
\end{picture}
\end{center}
\caption[]{ Generic Feynman graphs for invisible neutralino decays.}
\label{fig:graphs2}
\end{figure}
The set of Feynman diagrams involved in neutralino decays is indicated
in Figs.~\ref{fig:graphs}, \ref{fig:graphs1} and \ref{fig:graphs2}.
Most of the relevant couplings involved have been given in appendix B
of ref.~\cite{nulong} and the remaining ones will be included in
appendix B of the present paper. Even though these are sufficient for
our calculation of neutralino production and decay properties, it is
very useful to have approximate formulas for the neutralino couplings,
since this allows some qualitative understanding of the correlations
we are going to discuss.
To achieve this we make use of the expansions for the neutralino mass
matrix and also a corresponding one for the charginos as given in
\cite{Martin0nu}\footnote{Note that one has to reverse the sign of the
  $\epsilon_i$ in \cite{Martin0nu} to be consistent with our present
  notation.}.  For this purpose we will confine ourselves to the
tree-level neutralino/neutrino mass matrix and we refer to
\sect{sec:predictions} for a short discussion of the necessary changes
once the 1-loop corrections to the mass matrix are included.  However,
we have used exact numerical diagonalizations and loop effects in the
calculation of all resulting physical quantities presented in Secs. 3
and 4.

One class of decays which is important are those involving a
$W$-boson, either virtual or real. The $\chiz{1}$-$W^\pm$-$l_i$
couplings are approximatively given by:
\bea
\label{eq:coupWRChi}
O^{cnw}_{Ri1} &=& \frac{g h_E^{ii} v_D}{2 \mathrm{Det}_+} 
      \left[ \frac{g v_D N_{12} + M_2 N_{14}}{\mu} \epsilon_i \right.
     \nonumber  \\
     && \hspace{1.2cm} \left.
           + g \frac{ \left(2 \mu^2 + g^2 v_D v_U \right) N_{12}
                    + \left(\mu + M_2 \right) g v_U N_{14} }
                    {2 \mu \mathrm{Det}_+} \Lambda_i
             \right] \\
O^{cnw}_{Li1} &=& \frac{g \Lambda_i}{\sqrt{2}}
     \left[ - \frac{g' M_2 \mu }{2 \mathrm{Det}_0} N_{11} 
       + g \left( \frac{1}{\mathrm{Det}_+} 
                + \frac{M_1 \mu}{2 \mathrm{Det}_0} \right) N_{12} \right. \no
       & & \hspace{7mm} - \frac{v_U}{2} \left( 
                   \frac{g^2 M_1 + {g'}^2 M_2}{2 \mathrm{Det}_0}
                   + \frac{g^2}{\mu \mathrm{Det}_+} \right) N_{13} 
        + \left. \frac{v_D (g^2 M_1 + {g'}^2 M_2)}{4 \mathrm{Det}_0}
                 N_{14} \right] 
\label{eq:CoupWLChi}
\eea
Here $\mathrm{Det}_+$ and $\mathrm{Det}_0$ denote the determinant of
the MSSM chargino and neutralino mass matrix, respectively.  $N_{ij}$
are the elements of the mixing matrix which diagonalizes the MSSM
neutralino mass matrix.

For the coupling $Z$-$\chiz{1}$-$\nu_i$ we find
\begin{eqnarray}
\label{eq:coupZnu}
O^{nnz}_{L\chi^0_1 \nu_1} &=& O^{nnz}_{L\chi^0_1 \nu_2} = 0 \\
O^{nnz}_{L\chi^0_1 \nu_3} &=& \left(
        \frac{g \left(g M_1  N_{12}-  g' M_2 N_{11} \right) \mu}
             {4 \cos \theta_W \mathrm{Det}_0} 
      + \frac{g \left(g^2 M_1 + {g'}^2 M_2\right) v_D N_{14}}
             {4 \cos \theta_W \mathrm{Det}_0}  \right)
        |\vec{\Lambda}| \, \, . 
\end{eqnarray}
As already mentioned, the tree-level the states $\nu_1$ and $\nu_2$
are not well defined.  Therefore one has to consider the complete
1-loop mass matrix as it will be done in the numerical part in
sections 3 and 4. However, as one cannot detect single neutrino
flavours, in experiments one observes the decay of $\tilde \chi^0_1
\to X + \nu_i$ summing over all neutrinos $\nu_i$. Therefore, for the
$Z$-mediated decay the interesting quantity is $\sum_{i=1,3}
|{O^{nnz}_{L\chi^0_1 \nu_i}}|^2$ and, in contrast to the individual
$\chi^0_1 \to Z \, \nu_i$ decay rates, this only gets small radiative
corrections.

For the coupling $\chiz{1}$-$\nu_i$-($S^0_1 \simeq h^0$) we get
\begin{eqnarray}
\label{eq:coupSnuA}
O^{nnh}_{111} &=& E_{\chiz{1}} 
           \left( \sina \, c_2 \, c_4 \, c_6 
    \frac{- \epsilon_e \left( \Lambda_\mu^2 + \Lambda_\tau^2 \right)
          + \Lambda_e \left( \epsilon_\mu  \Lambda_\mu
                           + \epsilon_\tau \Lambda_\tau \right)}
         {\mu \sqrt{ \Lambda_e^2 +  \Lambda_\tau^2} |\vec{\Lambda}|}
        \right. \nonumber \\
 &&  \hspace{1cm} + \left.
    \frac{- s_2 \left( \Lambda_\mu^2 + \Lambda_\tau^2 \right)
          + \Lambda_e \left( s_4  \Lambda_\mu
                           + s_6 \Lambda_\tau \right)}
         { \sqrt{ \Lambda_e^2 +  \Lambda_\tau^2} |\vec{\Lambda}|} 
        \right) \\
O^{nnh}_{121} &=& E_{\chiz{1}} 
           \left( \sina \, c_2 \, c_4 \, c_6 
    \frac{ \epsilon_\tau \Lambda_\mu - \epsilon_\mu \Lambda_\tau }
         {\mu \sqrt{ \Lambda_e^2 +  \Lambda_\tau^2} } 
    + \frac{ s_6 \Lambda_\mu - s_4 \Lambda_\tau }
           { \sqrt{ \Lambda_e^2 +  \Lambda_\tau^2} } \right)  \\
O^{nnh}_{131} &=& E_{\chiz{1}}
        \left(\sina \, c_2 \, c_4 \, c_6
           \frac{(\vec{\epsilon}.\vec{\Lambda})}{|\vec{\Lambda}|}
         + \frac{(\vec{s}.\vec{\Lambda})}{|\vec{\Lambda}|} 
        \right)
       - D_{\chiz{1}} \, c_2 \, c_4 \, c_6 |\vec{\Lambda}|
\label{eq:coupSnu}
\end{eqnarray}
with
\begin{eqnarray}
\vec{s} &=& (s_2,s_4,s_6) \hspace{0.2cm}, \\
E_{\chiz{1}} &=& \frac{(g' N_{11} - g N_{12})}{2} \hspace{0.2cm}, \mathrm{and} \\
D_{\chiz{1}} &=&  \frac{\left( g^2 M_1 + {g'}^2 M_2 \right)
       \left[ \left( \cosa  \,v_D + \sina  \, v_U \right)
              \left(g' N_{11} - g N_{12} \right)
            + 2 \mu \left(\sina N_{13} - \cosa N_{14} \right) 
       \right]}{8 \mathrm{Det}_0} \no
\end{eqnarray}

The quantities $s_i$ and $c_i$ are parts of the mixing matrix for the neutral
scalars, which is discussed in \app{app:approximate}.  

For the couplings $\tilde d_{Li} - d_i - \nu_j$ one finds
\begin{eqnarray}
\label{eq:CoupSdDChi1}
 O^{dnL}_{Li1} &=& h^{ii}_D 
    \frac{- \epsilon_e \left( \Lambda_\mu^2 + \Lambda_\tau^2 \right)
          + \Lambda_e \left( \epsilon_\mu  \Lambda_\mu
                           + \epsilon_\tau \Lambda_\tau \right)}
         {\mu \sqrt{ \Lambda_e^2 +  \Lambda_\tau^2} |\vec{\Lambda}|} \\
 O^{dnL}_{Ri1} &=& 0 \\
 O^{dnL}_{Li2} &=& h^{ii}_D 
    \frac{ \epsilon_\tau \Lambda_\mu -  \epsilon_\mu \Lambda_\tau }
         {\mu \sqrt{ \Lambda_e^2 +  \Lambda_\tau^2} } \\
 O^{dnL}_{Ri2} &=& 0 \\
 O^{dnL}_{Li3} &=& h^{ii}_D \left(
    G_{\chiz{1}} |\vec{\Lambda}|
   - \frac{(\vec{\epsilon},\vec{\Lambda})}{\mu |\vec{\Lambda}|} \right) \\
 O^{dnL}_{Ri3} &=& H_{\chiz{1}} |\vec{\Lambda}|
\label{eq:CoupSdDChi2}
\end{eqnarray}
with $G_{\chiz{1}} = (g^2 M_1 + {g'}^2 M_2) v_U / (4 \mathrm{Det}_0)$
and $H_{\chiz{1}} = (3 g^2 M_1 + {g'}^2 M_2) \mu / (6 \sqrt{2}
\mathrm{Det}_0)$.  For the couplings $\tilde d_{Ri} - d_i - \nu_j$ one
finds that $O^{dnR}_{Rij} = O^{dnL}_{Lij}$ and
$O^{dnR}_{Lij} = O^{dnL}_{Rij}$ as above but
with $H_{\chiz{1}} \to {g'}^2 M_2 \mu / (3 \sqrt{2} \mathrm{Det}_0)$.

One can obtain the couplings between $\tilde l_{Li}$-$l_i$-$\nu_j$ by
replacing $h_D \to h_E$ and $H_{\chiz{1}} \to (g^2 M_1 + {g'}^2 M_2)
\mu / (2 \sqrt{2} \mathrm{Det}_0)$ in the above equations.  For the
case of $\tilde l_{Ri}$-$l_i$-$\nu_j$ one finds the couplings by replacing
$h_D \to h_E$ and $H_{\chiz{1}} \to {g'}^2 M_2 \mu / ( \sqrt{2}
\mathrm{Det}_0)$.

For the couplings $\tilde u_{Li} - u_i - \nu_j$ one finds
\begin{eqnarray}
  O^{unL}_{Li1} &=& O^{unL}_{Ri1} =
  O^{unL}_{Li2}  =  O^{unL}_{Ri2} = 0 \\
  O^{unL}_{Li3} &=& - h_U^{ii} I_{\chiz{1}} |\vec{\Lambda}| \\
 O^{unR}_{Li3} &=& J_{\chiz{1}} |\vec{\Lambda}|
\end{eqnarray}
with $I_{\chiz{1}} = (g^2 M_1 + {g'}^2 M_2) v_D / (2 \mathrm{Det}_0)$
and $J_{\chiz{1}} = (- 3 g^2 M_1 + {g'}^2 M_2) \mu / (6 \sqrt{2}
\mathrm{Det}_0)$.  For the couplings $\tilde u_{Ri}$-$u_i$-$\nu_j$ one
finds that $O^{unR}_{Rij} = O^{unL}_{Lij}$ and
$O^{unR}_{Lij} = O^{unL}_{Rij}$ as above but
with $J_{\chiz{1}} \to - \sqrt{2} {g'}^2 M_2 \mu / (3
\mathrm{Det}_0)$.

For the couplings $\tilde u_j - d_k - l_i$ one finds
\begin{eqnarray}
 \label{eq:char1}
 C^{\tilde u}_{Lk l_i} &=& h_D^{kk} R^{\tilde u}_{j1} \left( 
          \frac{\epsilon_i}{\mu}
        + \frac{g^2 v_U}{2 \mu} \frac{\Lambda_i}{ \mathrm{Det}_+}
                   \right)  \\
 C^{\tilde u}_{R l_i} &=& \frac{h_E^{ii} v_D}{\sqrt{2} \mathrm{Det}_+}
     \left\{ 
     \left( \frac{g^2 v_D R^{\tilde u}_{j1}}{\sqrt{2} }
           + h_U^{kk} M_2 R^{\tilde u}_{j2} \right) 
            \frac{\epsilon_i}{\mu} \right. \nonumber \\
    &&\hspace{16mm} + \left.
          \left[ \frac{g^2 \mu  R^{\tilde u}_{j1}}{\sqrt{2}}
                 \left(1 + \frac{g^2 v_D v_U}{2 \mu^2} \right)
                    + \frac{g^2 h_U^{kk} v_U R^{\tilde u}_{j2}}{2}
                 \left(1 + \frac{M_2}{\mu} \right)  
               \right] \frac{\Lambda_i}{\mathrm{Det}_+} \right\}
\end{eqnarray}

For the couplings $\tilde d_j - u_k - l_i$ one finds
\begin{eqnarray}
 C^{\tilde d}_{Lk l_i} &=& \frac{h_E^{ii} h_U^{kk} v_D
                  R^{\tilde d}_{j1}}{\sqrt{2}\mathrm{Det}_+}
        \left[ M_2 \frac{\epsilon_i}{\mu}
        + \frac{g^2 v_U }{2}
          \left( 1 + \frac{M_2}{\mu} \right)
          \frac{\Lambda_i}{ \mathrm{Det}_+}
                   \right]  \\
 C^{\tilde d}_{Rk l_i} &=& h_D^{kk} R^{\tilde d}_{j2} \frac{\epsilon_i}{\mu}
          + \left( \frac{g^2 R^{\tilde d}_{j1}}{\sqrt{2}} 
                 + \frac{g^2 h_D^{kk} v_U R^{\tilde d}_{j2}}{2 \mu} \right)
             \frac{\Lambda_i}{\mathrm{Det}_+}
 \label{eq:char4}
\end{eqnarray}
In \eq{eq:char1} - (\ref{eq:char4}) we have assumed that there is no
generation mixing between the squarks implying that $j=1,2$.

Data from reactor experiments~\cite{CHOOZ} indicate that the mixing
element $U_{e3}$ must be small~\cite{latestglobalanalysis}. This
implies that $|\Lambda_e| \ll |\Lambda_{2,3}|$.  In the limit
$\Lambda_e / \Lambda_{2,3}\to 0$ some of the above formulas simplify
to
\begin{eqnarray}
\label{eq:simple1}
O^{nnh}_{111} &=& - E_{\chiz{1}} 
         \left( \frac{ c_2 \, c_4 \, c_6 \, \sina \, \epsilon_e }{\mu } + s_2 
         \right) \\
O^{nnh}_{121} &=& E_{\chiz{1}} 
           \left( \sina \, c_2 \, c_4 \, c_6 
    \frac{ \epsilon_\tau \Lambda_\mu - \epsilon_\mu \Lambda_\tau }
         {\mu |\Lambda_\tau| } 
    + \frac{ s_6 \Lambda_\mu - s_4 \Lambda_\tau }{|\Lambda_\tau|} \right)  \\
O^{nnh}_{131} &=& E_{\chiz{1}}
        \left(\sina  \,c_2  \, c_4 \, c_6
    \frac{ \epsilon_\mu \Lambda_\mu + \epsilon_\tau \Lambda_\tau }
         {\mu \sqrt{\Lambda^2_2+\Lambda^2_3} } 
    + \frac{ s_4 \Lambda_\mu + s_6 \Lambda_\tau }{\sqrt{\Lambda^2_2+\Lambda^2_3}}
           \right) 
       - D_{\chiz{1}} \, c_2 \, c_4 \, c_6 \sqrt{\Lambda^2_2+\Lambda^2_3} \\
O^{dnL}_{Li\nu_1} &=& O^{dnR}_{Ri\nu_1}
                          = -  \frac{h_D^{ii} \epsilon_e}{\mu} \\
O^{dnL}_{Li\nu_2} &=& O^{dnR}_{Ri\nu_2}
  =  \frac{h_D^{ii} \left( \epsilon_\mu \Lambda_\tau- \epsilon_\tau \Lambda_\mu\right) }
          {\mu |\Lambda_\tau|} \\
O^{dnL}_{Li\nu_3} &=& O^{dnR}_{Ri\nu_3}
  =  h_D^{ii} \left( G_{\chiz{1}} \sqrt{\Lambda_\mu^2 + \Lambda^2_\tau}
         - \frac{\epsilon_\mu \Lambda_\mu+ \epsilon_\tau \Lambda_\tau}
          {\mu \sqrt{\Lambda_\mu^2 + \Lambda_\tau^2}} \right) 
\label{eq:simple6}
\end{eqnarray}
Later on we will also use the so-called sign-condition~\cite{nulong},
defined by
\bea
\frac{\epsilon_\mu}{\epsilon_\tau} \frac{\Lambda_\mu}{\Lambda_\tau} < 0 \, \, .
\eea 
Its origin can be traced back to the above \eq{eq:simple1} -
(\ref{eq:simple6}). Assuming $\epsilon_\mu \simeq \epsilon_\tau$ as
indicated by unification and $|\Lambda_\mu| \simeq |\Lambda_\tau|$ as
required by the atmospheric neutrino problem one sees easily from the
above equations that either the $\epsilon$ part\footnote{The $\Lambda$
  parts lead only to a renormalization of the heaviest neutrino state
  whereas the $\epsilon$ part gives mass to the lighter neutrinos.}
of the couplings to the second or the third neutrino state is very
small depending on the relative sign between $\Lambda_\mu$ and
$\Lambda_\tau$.  If $\Lambda_\mu \simeq -\Lambda_\tau$ one can show, after a
lengthy calculation \cite{MarcoFuture}, that the resulting effective
neutrino mixing matrix is given by
\bea
V_{\nu,\mathrm{loop}}=
\left(\begin{array}{ccc}
  \cos\theta_{12} &  -\sin\theta_{12} & 0 \\
  \sin\theta_{12} &  \cos\theta_{12}  & 0 \\
                0 & 0 &               1 
\end{array}\right) \times V_{\nu,\mathrm{tree}}
\eea
with nearly unchanged $\theta_{13}$ and $\theta_{23}$. In contrast, if
this sign condition is not fulfilled the $\theta_{13}$ and
$\theta_{23}$ angles get large corrections. One sees therefore that if
the sign condition is satisfied the atmospheric and solar neutrino
features decouple: the atmospheric is mainly tree-level physics, while
the solar neutrino anomaly is accounted for by genuine loop physics
in a simple factorizable way. Thus the sign condition is helpful to
get a better control on the parameters for the solar neutrino problem.

\section{Neutralino Production and Decays}
\label{sec:decays}

\begin{figure}
\setlength{\unitlength}{1mm}
\begin{picture}(150,90)
\put(-3,-50){\mbox{\epsfig{figure=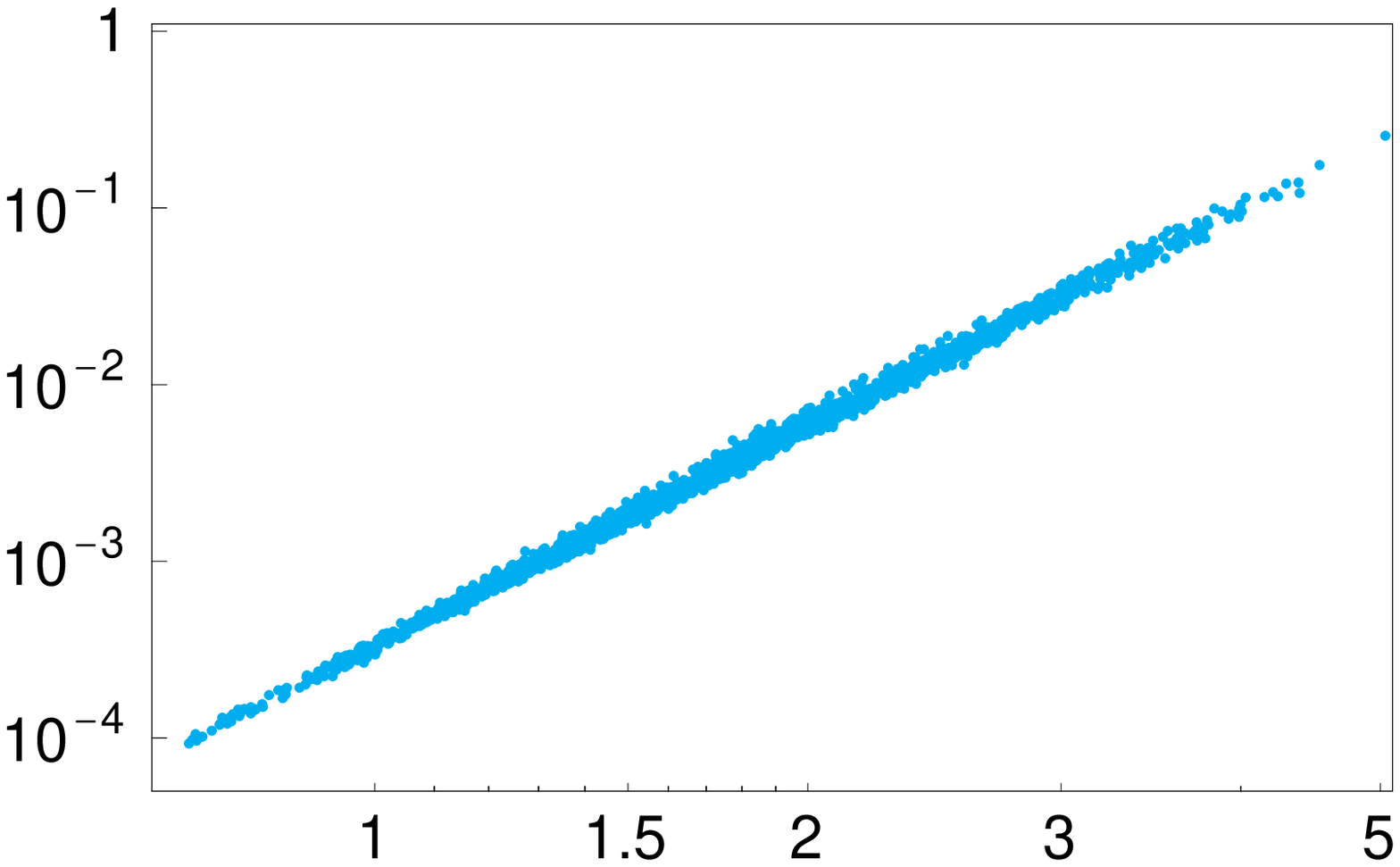,height=18.7cm,width=7.cm}}}
\put(-2,86){\makebox(0,0)[bl]{{ a)}}}
\put(3,85){\makebox(0,0)[bl]{{\small $\Delta m^2_{atm}$}}}
\put(77,-3){\makebox(0,0)[br]{
             {$10^5 |{\vec \Lambda}|/(\sqrt{M_2} \mu)$ $[GeV]$}}}
\put(82,0){\mbox{\epsfig{figure=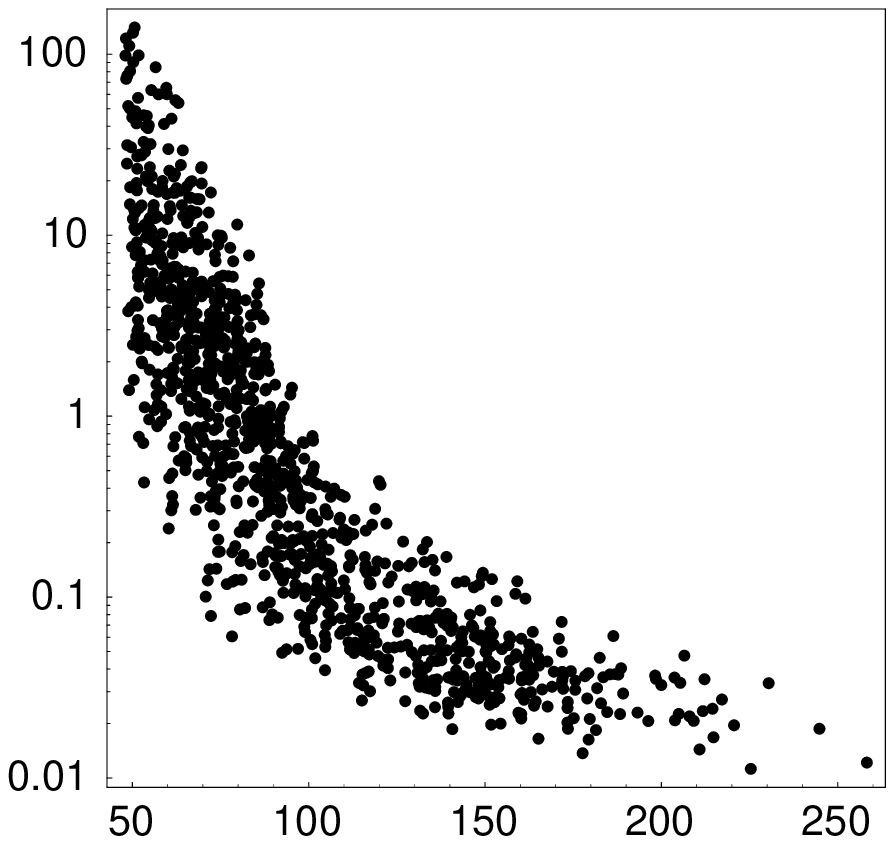,height=8.7cm,width=7.cm}}}
\put(82,87){\makebox(0,0)[bl]{{ b)}}}
\put(87,85){\makebox(0,0)[bl]{{\small $L(\chiz{1})$~[cm]}}}
\put(158,-3){\makebox(0,0)[br]{{ $m_{\chiz{1}}$~[GeV]}}}
\end{picture}
\caption[]{ a) $\Delta m^2_{atm}$ and  b) neutralino decay length.}
\label{fig:TestAtmMass}
\end{figure}

\begin{figure}
\setlength{\unitlength}{1mm}
\begin{center}
\begin{picture}(80,80)
\put(0,0){\mbox{\epsfig{figure=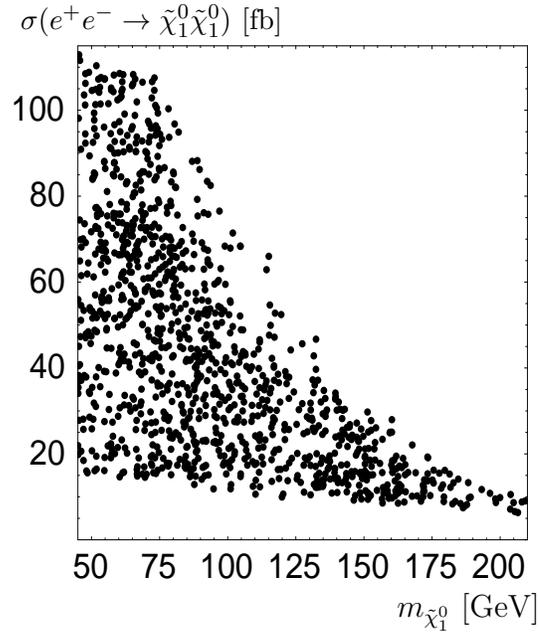,height=7.7cm,width=7.cm}}}
\put(2,76){\makebox(0,0)[bl]
            {{\small $\sigma(e^+ e^- \to \chiz{1} \chiz{1})$~[fb]}}}
\put(71,-3){\makebox(0,0)[br]{{$\mchiz{1}$~[GeV]}}}
\end{picture}
\end{center}
\caption[]{Production cross section for the process
$\sigma(e^+ e^- \to \chiz{1} \chiz{1})$ as a function of $\mchiz{1}$ at a
Linear Collider with 1~TeV c.m.s energy. ISR-corrections are included.}
\label{fig:Prod1TeV}
\end{figure}

In this section we will discuss the production and the decay modes of
the lightest neutralino $\chiz{1}$.  In order to reduce the numbers of
parameters we have performed the calculations in the framework of a
minimal SUGRA version of bilinearly \rp SUSY model. Unless noted
otherwise the parameters have been varied in the following ranges:
$M_2$ and $|\mu|$ from 0 to 1 TeV, $m_0$ [0.2 TeV, 1.0 TeV], $A_0/m_0$
and $B_0/m_0$ [-3,3] and $\tan\beta$ [2.5,10], and for the \rp
parameters, $|\Lambda_\mu/\sqrt{\Lambda_e^2+\Lambda^2_\tau}|= 0.4-2$,
$\epsilon_\mu/\epsilon_\tau = 0.8-1.25$, $|\Lambda_e/\Lambda_\tau|=
0.025-2$, $\epsilon_e/\epsilon_\tau = 0.015-2$ and $|\Lambda| =
0.05-0.2$ GeV$^2$. They were subsequently tested for consistency with
the minimization (tadpole) conditions of the Higgs potential as well
as for phenomenological constraints from supersymmetric particle
searches. Moreover, they were checked to provide a solution to both
solar and atmospheric neutrino problems. For the case of the solar
neutrino anomaly we have accepted points which give either one of the
large mixing angle solutions or the small mixing angle MSW solution.

We have seen in \eq{mnutree} that the atmospheric scale is
proportional $|\vec{\Lambda}|^2/ \mathrm{Det}(\mchiz{})$. As has been
shown in \cite{nushort,nulong} this statement remains valid after
inclusion of 1-loop corrections provided that
$|\vec{\epsilon}|^2/|\vec{\Lambda}| < 1$ implying that 1-loop
corrections to the heaviest neutrino mass remain small. As we have
seen in section (\ref{sec:approx}), most of the couplings are
proportional to $|\vec{\Lambda}|/ \sqrt{\mathrm{Det}(\mchiz{}})$
and/or $\epsilon_i / \mu$.
Although $|\vec{\Lambda}|/(\sqrt{M_2}\mu)$ has to be small in order to
account for the atmospheric mass scale (see \figz{fig:TestAtmMass}{a})
the previously discussed couplings are still large enough so that the
neutralino decays inside the detector, as can be seen in
\figz{fig:TestAtmMass}{b}.

In \fig{fig:Prod1TeV} we show the cross section $\sigma(e^+ e^- \to
\chiz{1} \chiz{1})$ in fb for $\sqrt{s} = 1$~TeV.  Assuming now that
an integrated luminosity of 1000~fb$^{-1}$ per year can be achieved at
a future linear collider (see \cite{tesla,blair00} and references
therein) this implies that between $10^4$ to $10^5$ neutralino pairs
can be directly produced per year. Due to the smallness of the
R-parity violating couplings, most of the SUSY particles will decay
according to the MSSM scheme implying that there will be many more
neutralinos to study, namely from direct production as well as
resulting from cascade decays of heavier SUSY particles. From this
point of view the measurement of branching ratios as small as
$10^{-5}$ should be feasible. As we will see in what follows this
might be required in order to establish some of the correlations
between neutrino mixing angles and the resulting neutralino decay
observables, which is a characteristic feature of this class of models.
 
In this model the neutralino can decay in the following ways
\bea
\chiz{1} &\to& \nu_i \, \nu_j \, \nu_k \\
         &\to& \nu_i \, q \, \bar{q} \\
         &\to& \nu_i \, l^+_j \, l^-_k \\
         &\to& l^\pm_i \, q \, \bar{q}' \\
         &\to& \nu_i \, \gamma
\eea
In the following we will discuss these possibilities in detail except
$\chiz{1} \to \nu_i \, \gamma$ because its branching ratio is always
below $10^{-7}$.

In the following discussion we have always computed the complete
three-body decay widths even in cases where $\mchiz{1}$ has been
larger than one of the exchanged particle masses, so that two-body
channels are open. This has turned out to be necessary because there
are parameter combinations where the couplings to the lightest
exchanged particle are $O(10)$ smaller than the coupling to one of the
heavier particles, implying that the graph containing the heavy
particle cannot be neglected with respect to the lighter particle
contribution.  An example is the case of $Z$-boson and
$S^0_1$-mediated gaugino-like neutralino decays discussed later on.
Here $S^0_1$ denotes the lightest neutral scalar.
In addition we want to be sure not to miss possibly important
interference effects as there are several graphs which contribute to a
given process. A typical example is the process $\chiz{1} \to \nu_i \,
l^-_j \, l^+_k$ where 26 contributions exist, as can be seen from the
generic diagrams  shown in \fig{fig:graphs}.

\begin{figure}
\setlength{\unitlength}{1mm}
\begin{center}
\begin{picture}(80,80)
\put(0,-1){\mbox{\epsfig{
     figure=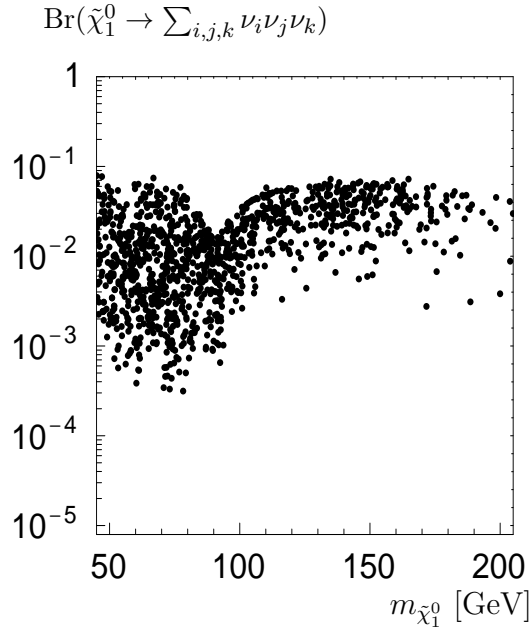,height=7.7cm,width=7.cm}}}
\put(5,75){\makebox(0,0)[bl]{{\small
        Br(${\tilde \chi}^0_1 \to \sum_{i,j,k} \nu_i \nu_j \nu_k$)}}}
\put(70,-3){\makebox(0,0)[br]{{$m_{{\tilde \chi}^0_1}$~[GeV]}}}
\end{picture}
\end{center}
\caption[]{Invisible neutralino branching ratio summing over all neutrinos.}
\label{fig:chiInv}
\end{figure}

The first important question to be answered is how large the invisible
neutralino decay modes to neutrinos can be. This is important to
ensure that sufficient many neutralino decays can be observed. As can
be seen from \fig{fig:chiInv} the invisible branching ratio never
exceeds 10\%.  
The main reason for this behaviour can be found in the fact that 
for the SUGRA motivated scenario under consideration the couplings 
of the lightest neutralino to the Z-boson are suppressed.
This and the comparison with other couplings will be discussed in some
detail later on.

\begin{figure}
\setlength{\unitlength}{1mm}
\begin{picture}(150,75)
\put(0,0){\mbox{\epsfig{
          figure=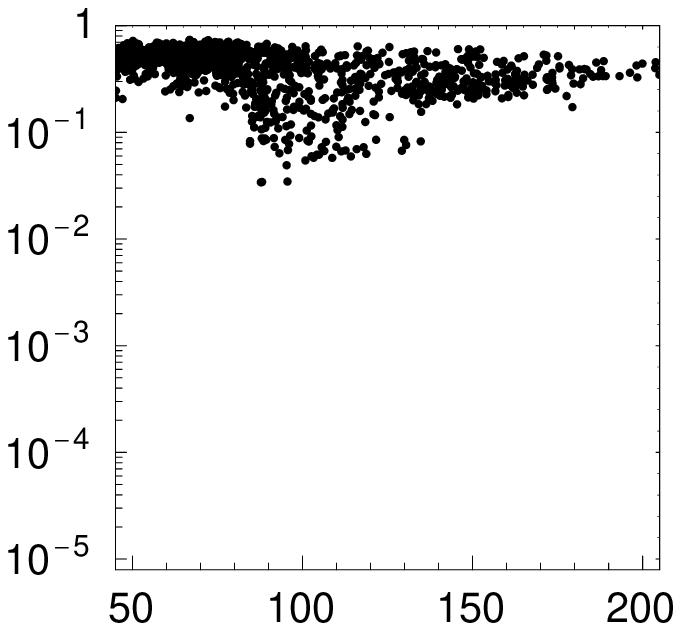,height=7.cm,width=5.cm}}}
\put(0,68){\makebox(0,0)[bl]{{\small
                  a)   Br(${\tilde \chi}^0_1 \to b \bar{b} \sum_i \nu_i$)}}}
\put(50,-3){\makebox(0,0)[br]{{$m_{{\tilde \chi}^0_1}$~[GeV]}}}
\put(54,0){\mbox{\epsfig{
        figure=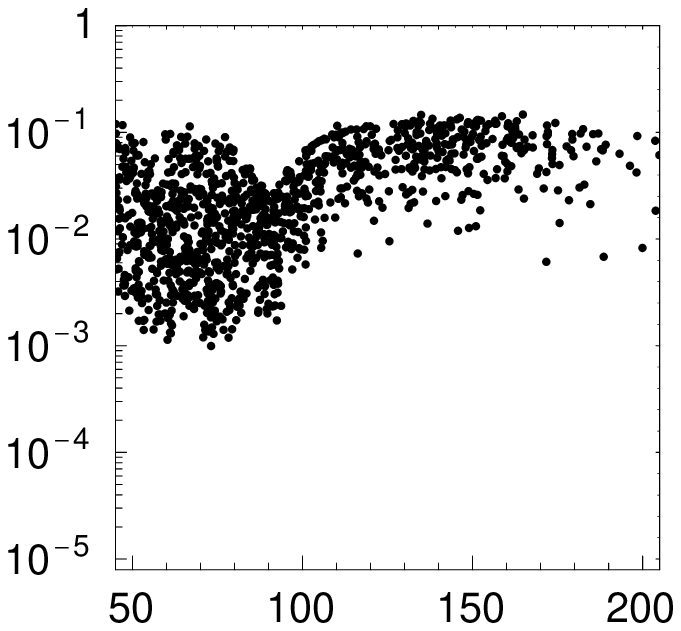,height=7.cm,width=5.cm}}}
\put(54,68){\makebox(0,0)[bl]{{\small
     b) Br(${\tilde \chi}^0_1 \to  \sum_{q=u,d,s} q \bar{q}  \sum_i \nu_i$)}}}
\put(104,-3){\makebox(0,0)[br]{{$m_{{\tilde \chi}^0_1}$~[GeV]}}}
\put(108,0){\mbox{\epsfig{
        figure=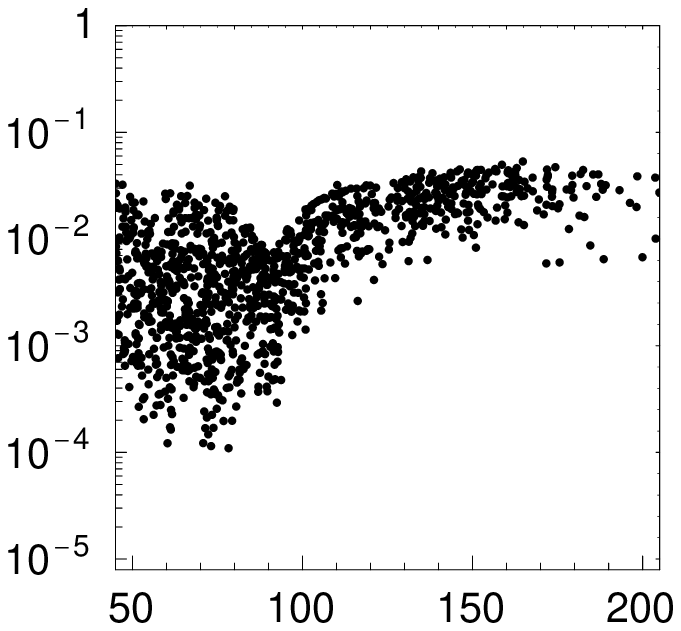, height=7.cm,width=5.cm}}}
\put(108,68){\makebox(0,0)[bl]{{\small
                 c)  Br(${\tilde \chi}^0_1 \to c \bar{c} \sum_i \nu_i$)}}}
\put(158,-3){\makebox(0,0)[br]{{$m_{{\tilde \chi}^0_1}$~[GeV]}}}
\end{picture}
\caption[]{Neutralino branching ratios for the decays into $q \bar{q} \nu_i$
           final states summing over all neutrinos. }
\label{fig:chiQQnu}
\end{figure}
\begin{figure}
\setlength{\unitlength}{1mm}
\begin{picture}(150,75)
\put(0,0){\mbox{\epsfig{
          figure=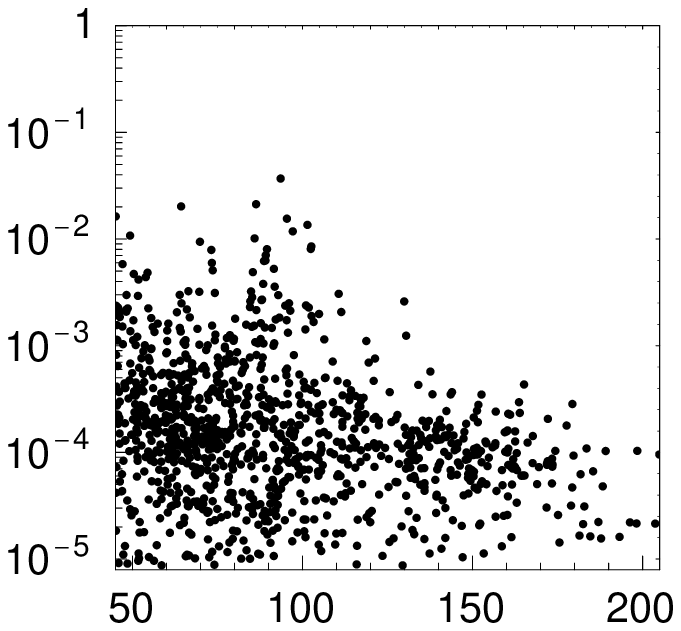,height=7.cm,width=5.cm}}}
\put(0,68){\makebox(0,0)[bl]{{\small
                  a)   Br(${\tilde \chi}^0_1 \to e^\pm \sum \bar{q} q'$)}}}
\put(50,-3){\makebox(0,0)[br]{{$m_{{\tilde \chi}^0_1}$~[GeV]}}}
\put(54,0){\mbox{\epsfig{
        figure=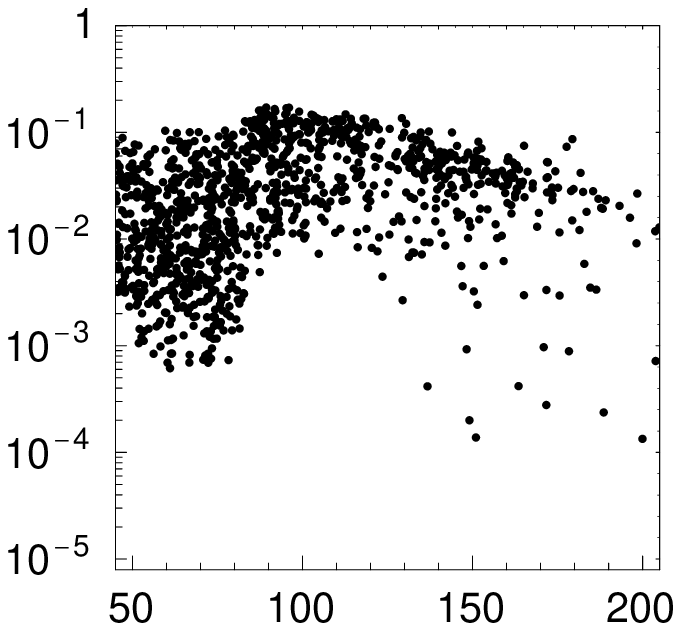,height=7.cm,width=5.cm}}}
\put(54,68){\makebox(0,0)[bl]{{\small
     b) Br(${\tilde \chi}^0_1 \to \mu^\pm \sum \bar{q} q'$)}}}
\put(104,-3){\makebox(0,0)[br]{{$m_{{\tilde \chi}^0_1}$~[GeV]}}}
\put(108,0){\mbox{\epsfig{
        figure=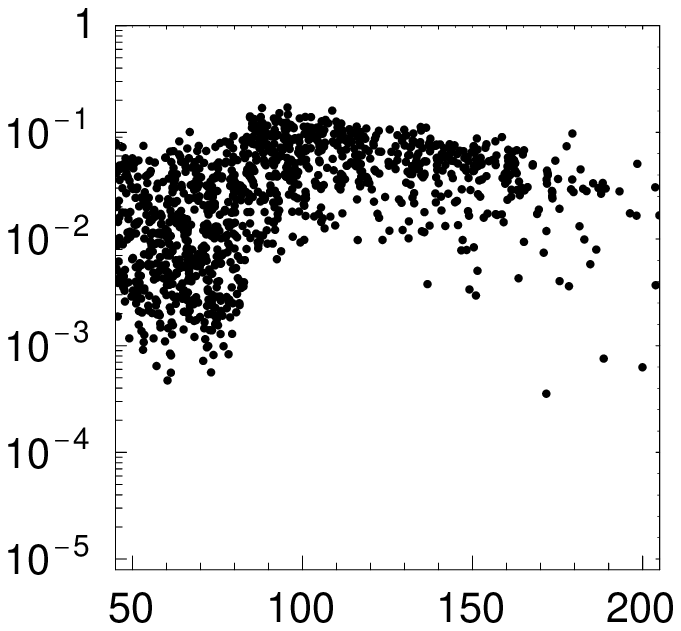, height=7.cm,width=5.cm}}}
\put(108,68){\makebox(0,0)[bl]{{\small
                 c)  Br(${\tilde \chi}^0_1 \to \tau^\pm \sum \bar{q} q'$)}}}
\put(158,-3){\makebox(0,0)[br]{{$m_{{\tilde \chi}^0_1}$~[GeV]}}}
\end{picture}
\caption[]{Neutralino branching ratios for the decays into
           $l^\pm q'\bar{q}$ final states summing over all $q' \bar{q}$
           combinations.}
\label{fig:chiLQQp}
\end{figure}
The mainly ``visible'' nature of the lightest neutralino decay,
together with the short neutralino decay path discussed above,
suggests the observability of neutralino-decay-induced events at
collider experiments and this should stimulate dedicated detector
studies.

In \fig{fig:chiQQnu} we show the branching ratios for the decays into
$q \bar{q} \nu_i$. Here we single out the $b$-quark
(\figz{fig:chiQQnu}{a}) and the $c$-quark (\figz{fig:chiQQnu}{c})
because in these cases flavour detection is possible. One can clearly
see that for $\mchiz{1} \lsim 1.1 \, m_W$ the decay into $b \bar{b}
\nu_i$ can be the dominant one. The reason is that the scalar
contributions stemming from $S^0_j$, $P^0_j$ and/or $\tilde b_k$ can
be rather large. This can be understood with the help of
\eq{eq:CoupSdDChi1}-(\ref{eq:CoupSdDChi2}) where terms proportional to
$h_D \epsilon_i / \mu$ appear. This kind of terms is absent in the
corresponding couplings for the u-type squarks implying that the
branching ratio for $c \bar{c} \nu_i$ is rather small as can be seen
in \figz{fig:chiQQnu}{c}. One can see in \figz{fig:chiQQnu}{b} and
\figz{fig:chiQQnu}{c} a pronounced 'hole' around 80-100 GeV. It occurs
because for $\mchiz{1} > m_W$ the $W$ becomes on-shell implying a
reduction for these decays. This is compensated as the $Z$ becomes
on-shell.

\begin{figure}
\setlength{\unitlength}{1mm}
\begin{picture}(150,160)
\put(0,79){\mbox{\epsfig{
             figure=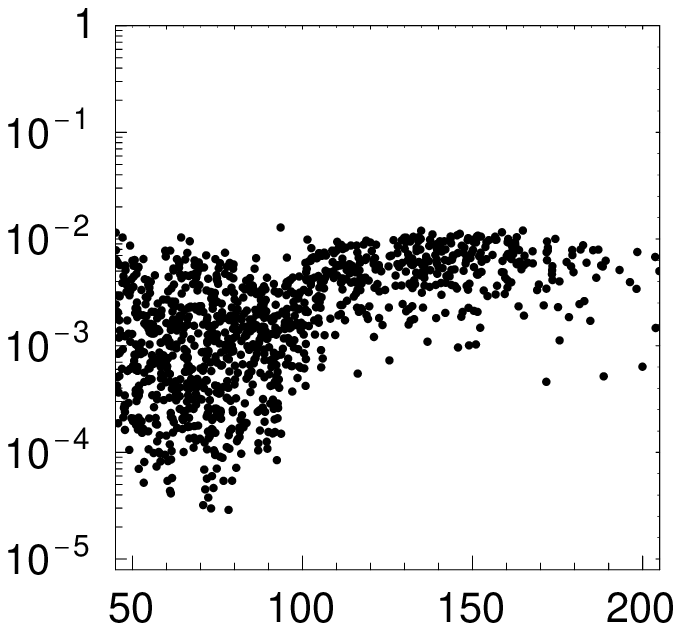,height=7.cm,width=5.cm}}}
\put(0,148){\makebox(0,0)[bl]{{\small
                  a) Br(${\tilde \chi}^0_1 \to e^- e^+ \sum_i \nu_i$)}}}
\put(50,77){\makebox(0,0)[br]{{$m_{{\tilde \chi}^0_1}$~[GeV]}}}
\put(54,79){\mbox{\epsfig{
         figure=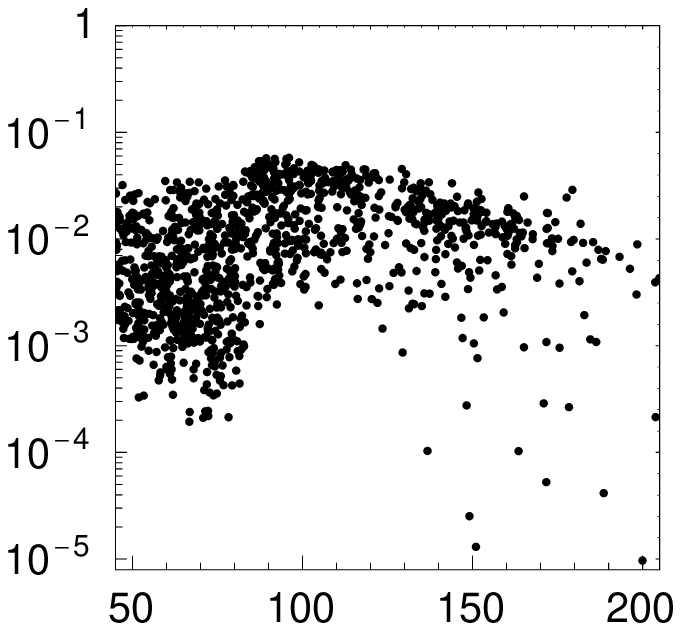,height=7cm,width=5.cm}}}
\put(54,148){\makebox(0,0)[bl]{{\small
                 b) Br(${\tilde \chi}^0_1 \to e^\pm \mu^\mp \sum_i \nu_i$)}}}
\put(104,77){\makebox(0,0)[br]{{$m_{{\tilde \chi}^0_1}$~[GeV]}}}
\put(108,79){\mbox{\epsfig{
         figure=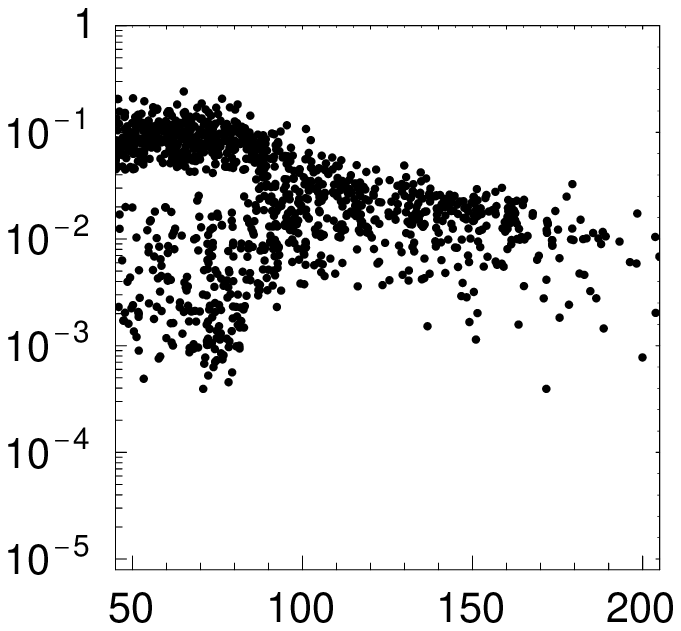,height=7cm,width=5.cm}}}
\put(108,148){\makebox(0,0)[bl]{{\small
                 c) Br(${\tilde \chi}^0_1 \to e^\pm \tau^\mp \sum_i \nu_i$)}}}
\put(158,77){\makebox(0,0)[br]{{$m_{{\tilde \chi}^0_1}$~[GeV]}}}
\put(0,0){\mbox{\epsfig{
          figure=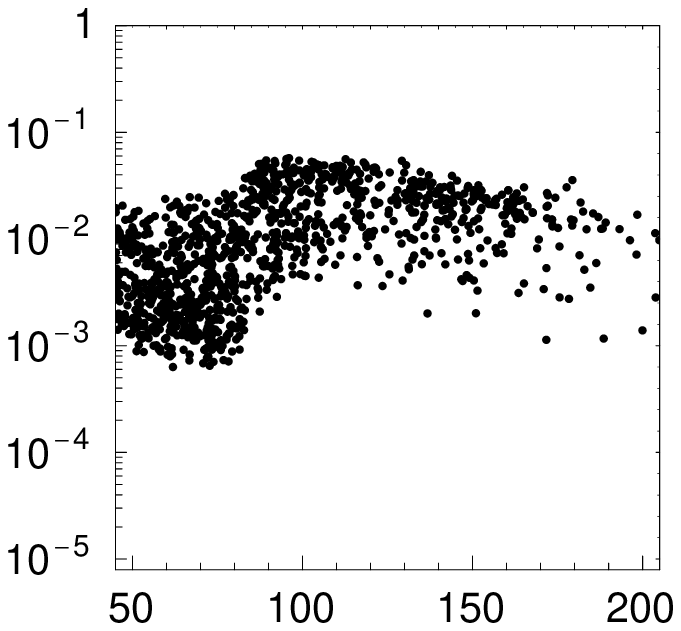,height=7.cm,width=5.cm}}}
\put(0,68){\makebox(0,0)[bl]{{\small
                  d)   Br(${\tilde \chi}^0_1 \to  \mu^- \mu^+ \sum_i \nu_i$)}}}
\put(50,-3){\makebox(0,0)[br]{{$m_{{\tilde \chi}^0_1}$~[GeV]}}}
\put(54,0){\mbox{\epsfig{
        figure=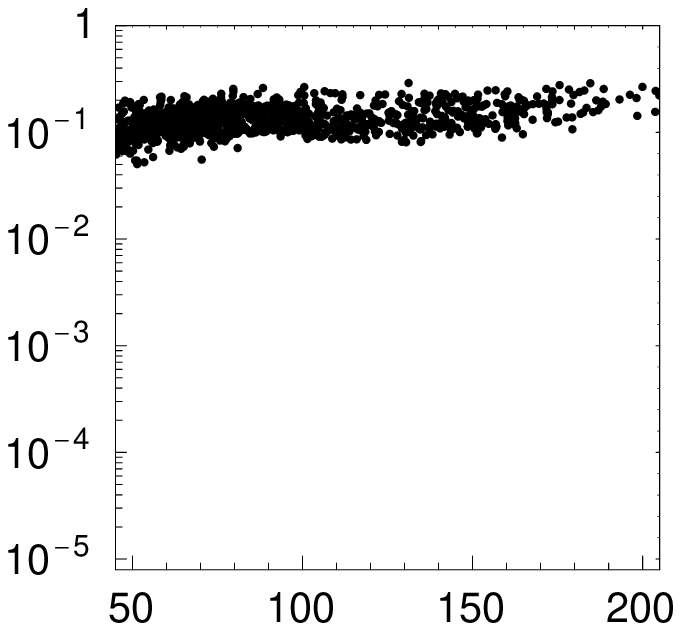,height=7.cm,width=5.cm}}}
\put(54,68){\makebox(0,0)[bl]{{\small
     e) Br(${\tilde \chi}^0_1 \to  \mu^\pm \tau^\mp \sum_i \nu_i$)}}}
\put(104,-3){\makebox(0,0)[br]{{$m_{{\tilde \chi}^0_1}$~[GeV]}}}
\put(108,0){\mbox{\epsfig{
        figure=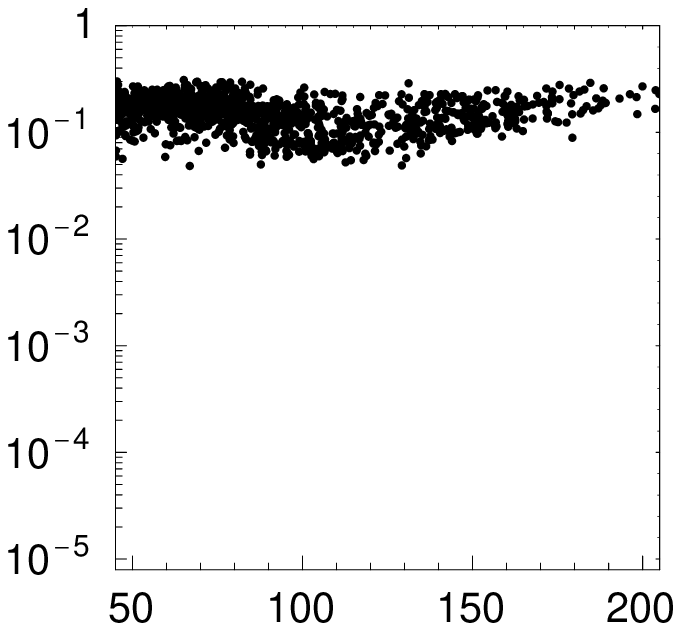, height=7.cm,width=5.cm}}}
\put(108,68){\makebox(0,0)[bl]{{\small
           f)  Br(${\tilde \chi}^0_1 \to  \tau^- \tau^+ \sum_i \nu_i$)}}}
\put(158,-3){\makebox(0,0)[br]{{$m_{{\tilde \chi}^0_1}$~[GeV]}}}
\end{picture}
\caption[]{Neutralino branching ratios for the decays into various lepton
           final states summing over all neutrinos.}
\label{fig:chiLLnu}
\end{figure}

The semi-leptonic branching ratios into charged leptons are shown in
\fig{fig:chiLQQp}. The decays into $\mu$ and $\tau$ are particularly
important because, as we will see in \sect{sec:correlations}, they
will give a measure of the atmospheric neutrino angle. Note that these
branching ratios are larger than $10^{-4}$ and in most cases larger
than $10^{-3}$, implying that there should be sufficient statistics
for investigations. In case of the $e$ final state it might happen
that one can only give an upper bound on this branching ratio.  This
is just a result of the reactor neutrino bound~\cite{CHOOZ}.
Note that due to the Majorana nature of the neutralino one expects in
large regions of the parameter space several events with same sign
di-leptons and four jets.

In \fig{fig:chiLLnu} the fully leptonic branching ratios are shown.
One can clearly see a difference between the branching ratios into
channels containing different charged leptons of the same flavour,
i.e.  $\tau^- \tau^+$ versus $\mu^- \mu^+$ and $e^- e^+$.
This difference is due to the importance of the $S^0_1$ state which
corresponds mainly to the lightest Higgs boson $h^0$ of the MSSM. We
have found that for gaugino-like $\chiz{1}$ the \rp couplings $S^0_1 -
\chiz{1} - \nu_i$ are in general larger than the corresponding $Z^0 -
\chiz{1} - \nu_i$ couplings.  This can be understood by inspecting the
formulas given in \eq{eq:coupZnu} - (\ref{eq:coupSnu}) in
\sect{sec:approx}, in particular the parts proportional to
$\epsilon_k$ in \eq{eq:coupSnuA} -- (\ref{eq:coupSnu}). Other reasons
for having ``non-universal'' $\tau^- \tau^+$, $\mu^- \mu^+$ and $e^-
e^+$ couplings are the graphs containing $W$ or charged sleptons as
exchanged particle (see \fig{fig:graphs}). From \eq{eq:coupWRChi} one
can see that the coupling $O^{cnw}_{Ri1}$ is proportional to
$h_E^{ii}$ implying that they only play a role if a $\tau$ is present
in the final state.

Notice also that the largeness of the branching ratios for neutralino
decays into lepton-flavour-violating channels can be simply understood
from the importance of $W^\pm$ and $S^\pm_n$ contributions present in
\fig{fig:graphs} \footnote{The charged scalars are a mixture of the
  charged Higgs bosons and the charged sleptons, and in particular the
  later are the important ones}.

\section{Probing Neutrino mixing via Neutralino Decays}
\label{sec:correlations}

In this section we will demonstrate that neutralino decay branching
ratios are strongly correlated with neutrino mixing angles. We will
consider two cases: 1) The situation before supersymmetry is
discovered. In this case we demonstrate that neutrino physics implies
predictions for neutralino decays which will be tested at future
colliders.   
2) The situation when the spectrum is known to the 1\% level or better
as could, for example, be achieved at a future linear collider
\cite{tesla,Martyn:1999tc}. In this case our model allows for several
consistency checks between neutrino physics (probed by underground and
reactor experiments~\cite{SuperK-atmos,solar,CHOOZ}) and neutralino
physics.
Moreover, some neutralino decay observables are sensitive to which of
the possible solutions to the solar neutrino problem is the one
realized, i.e. they can discriminate large angles solutions from the
small angle MSW solution.
%

\subsection{Before SUSY is discovered}
\label{sec:predictions}
 
\begin{figure}
\setlength{\unitlength}{1mm}
\begin{center}
\begin{picture}(80,80)
\put(0,1){\mbox{\epsfig{figure=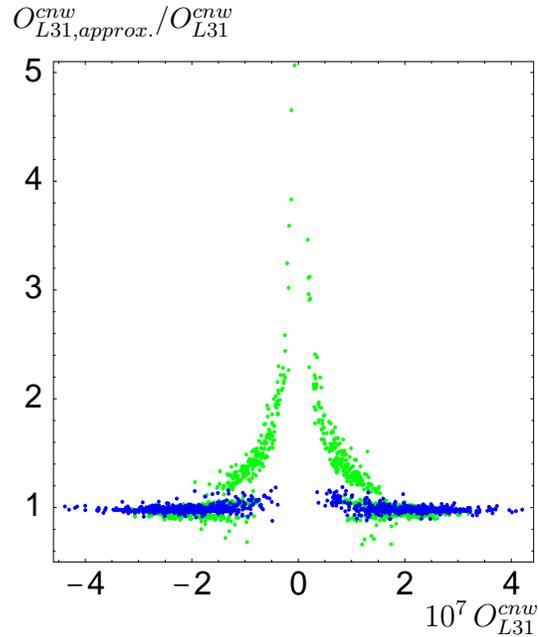,height=7.3cm,width=7.cm}}}
\put(0,76){\makebox(0,0)[bl]{{\small
        $O^{cnw}_{L31,approx.}/O^{cnw}_{L31}$}}}
\put(70,-3){\makebox(0,0)[br]{{$10^7 \, O^{cnw}_{L31}$}}}
\end{picture}
\end{center}
\caption[]{Approximated coupling $O^{cnw}_{L31,approx.}$ using formula 
           \eq{eq:CoupWLChi} 
           divided by the exact calculated coupling as a function of
           the exact calculated coupling. The bright (dark) points are for 
           $\mu > (<) 0$.}
\label{fig:TestCoupling}
\end{figure}

Let us first consider the situation before SUSY is discovered.  Before
working out the predictions for neutralino decays we would like to
point out a fact concerning the 1-loop corrected neutrino/neutralino
mass matrix.  It has been noticed in \cite{nulong} that the sign of
the $\mu$ parameter determines to some extent how large the absolute
radiative corrections are (see Fig.~5 of \cite{nulong})~\footnote{The
  important information is the relative sign between $\mu$ and the
  gaugino mass parameters $M_{1,2}$. Since in \cite{nulong} as well as
  here we assume that $M_{1,2}>0$ then the absolute sign of $\mu$
  becomes relevant. }.  The reason is that depending on this sign the
interference between the 1-loop graphs containing gauginos and the
1-loop graphs containing Higgsinos is constructive or destructive.

This fact has of course implications on whether the approximate
couplings presented in \sect{sec:approx} remain valid after the 1-loop
corrections are taken into account~\footnote{Of course the couplings
  involving $\nu_1$ and/or $\nu_2$ are exceptional ones, as the angle
  between these states is only meaningful after performing 1-loop
  corrections are included.}.  A typical example is shown in
\fig{fig:TestCoupling} where the approximated coupling
$O^{cnw}_{L31,approx.}$ divided by the coupling $O^{cnw}_{L31}$ as a
function of $O^{cnw}_{L31}$. One clearly sees that for $\mu <0$ the
tree-level result~\cite{acceltestnew} is a good approximation to
within 20\%, but for $\mu >0$ it can be off by a factor up to 5 in
some extreme cases where a constructive interference between gaugino
and Higgsino loops takes place.
We have checked that the same is true for the other couplings
involving either the charged leptons and/or $\nu_3$.

\begin{figure}
\setlength{\unitlength}{1mm}
\begin{picture}(150,80)
\put(0,1){\mbox{\epsfig{
               figure=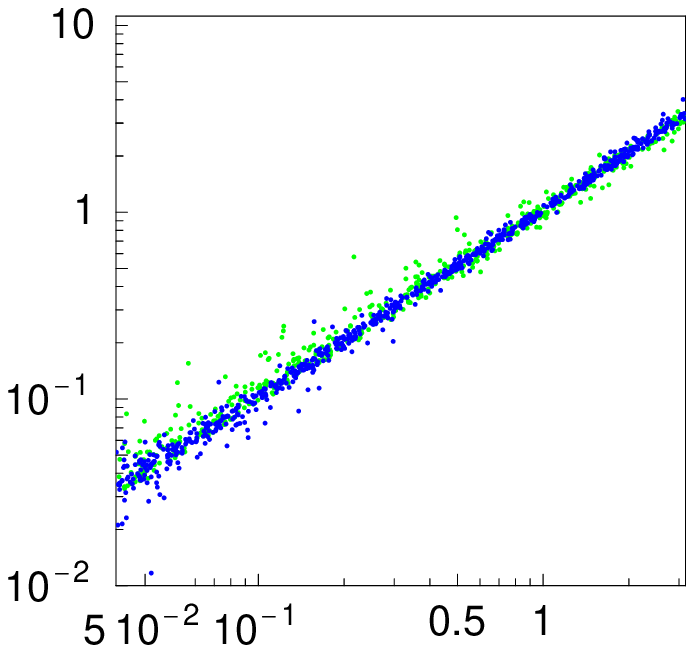,height=7.3cm,width=7.cm}}}
\put(0,75){\makebox(0,0)[bl]{{\small a) Br($\mu q q'$)/Br($\tau q q'$)}}}
\put(70,-3){\makebox(0,0)[br]{{$\tan^2(\theta_{atm})$}}}
\put(88,1){\mbox{\epsfig{figure=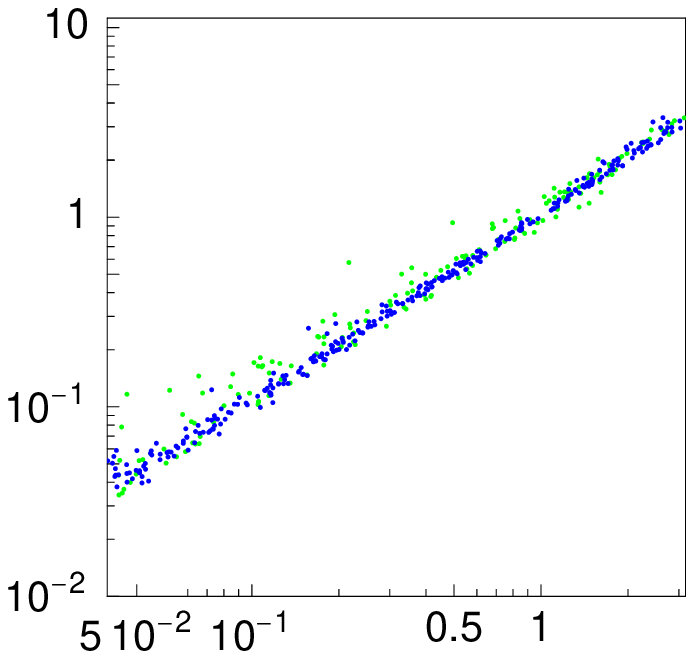,
                               height=7.3cm,width=7.cm}}}
\put(85,75){\makebox(0,0)[bl]{{\small b) Br($\mu q q'$)/Br($\tau q q'$)}}}
\put(158,-3){\makebox(0,0)[br]{{$\tan^2(\theta_{atm})$}}}
\end{picture}
\caption[]{Testing the atmospheric angle.
           In case of the dark (bright) points  $\mu<(>)0$. 
           In the second figure we have taken 
           only those points with $|\sin 2 \theta_{\tilde b}| > 0.1$.}
\label{fig:TestAngleAtm}
\end{figure}
\begin{figure}
\setlength{\unitlength}{1mm}
\begin{picture}(150,80)
\put(0,1){\mbox{\epsfig{
               figure=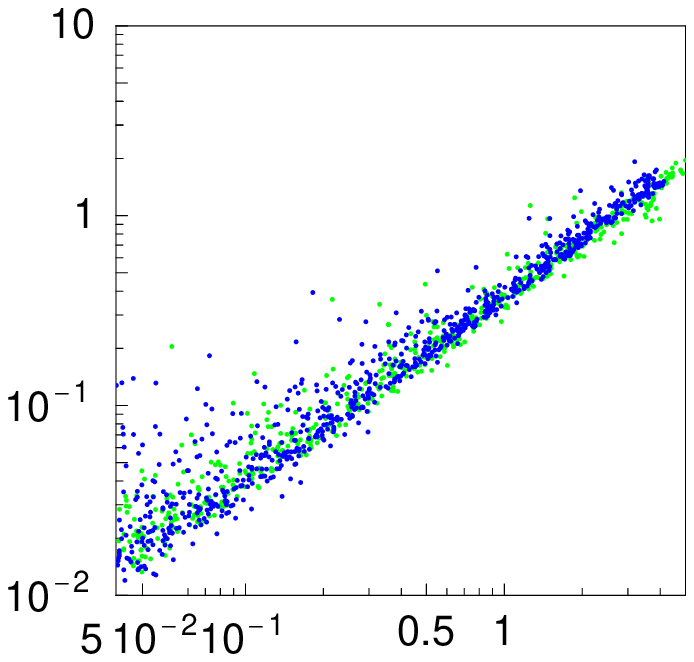,height=7.3cm,width=7.cm}}}
\put(0,75){\makebox(0,0)[bl]{{\small
        a) Br($e^\pm \mu^\mp \sum_i \nu_i$)/Br($\tau q q'$)}}}
\put(70,-3){\makebox(0,0)[br]{{$\tan^2(\theta_{atm})$}}}
\put(88,1){\mbox{\epsfig{figure=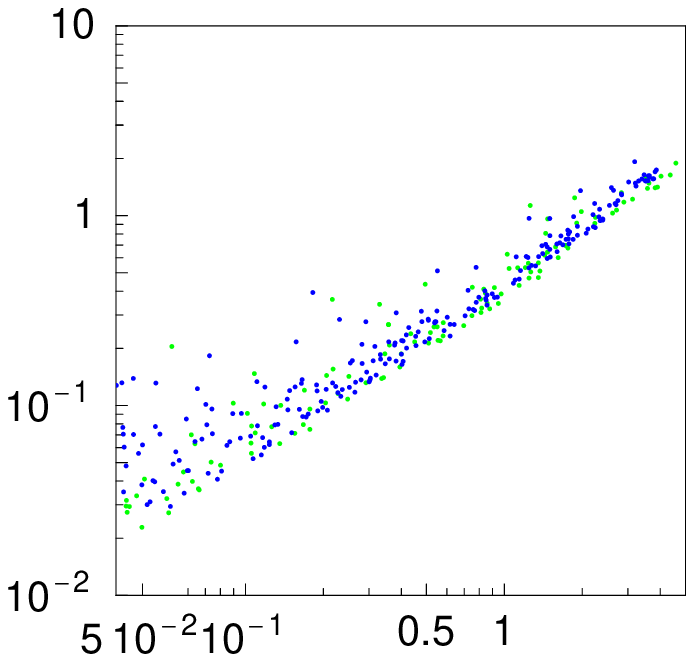,
                               height=7.3cm,width=7.cm}}}
\put(85,75){\makebox(0,0)[bl]{{\small
             b) Br($e^\pm \mu^\mp \sum_i \nu_i$)/Br($\tau q q'$)}}}
\put(158,-3){\makebox(0,0)[br]{{$\tan^2(\theta_{atm})$}}}
\end{picture}
\caption[]{Testing the atmospheric angle.
           In case of the dark (bright) points  $\mu<(>)0$. 
           In the second figure we have taken 
           only those points with $|\sin 2 \theta_{\tilde b}| > 0.1$.}
\label{fig:TestAngleAtmA}
\end{figure}
\begin{figure}
\setlength{\unitlength}{1mm}
\begin{picture}(150,80)
\put(0,1){\mbox{\epsfig{
               figure=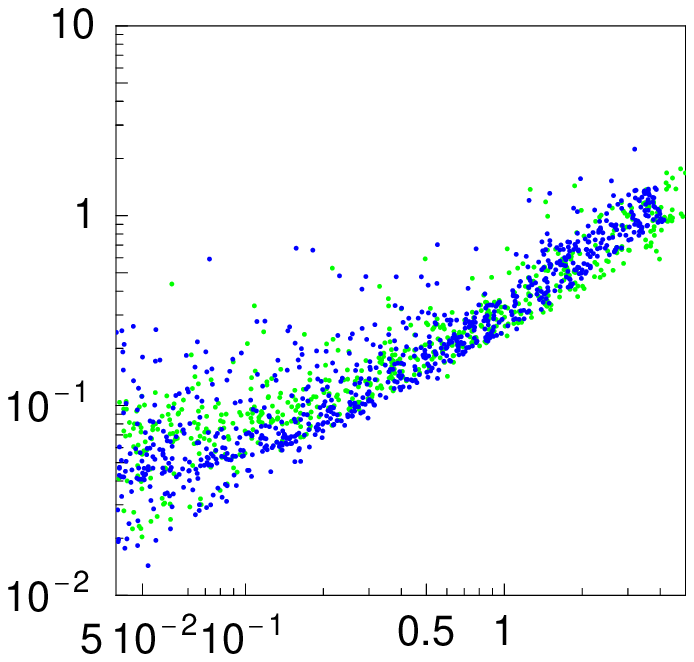,height=7.3cm,width=7.cm}}}
\put(0,75){\makebox(0,0)[bl]{{\small
        a) Br($\mu^- \mu^+ \sum_i \nu_i$)/Br($\tau q q'$)}}}
\put(70,-3){\makebox(0,0)[br]{{$\tan^2(\theta_{atm})$}}}
\put(88,1){\mbox{\epsfig{figure=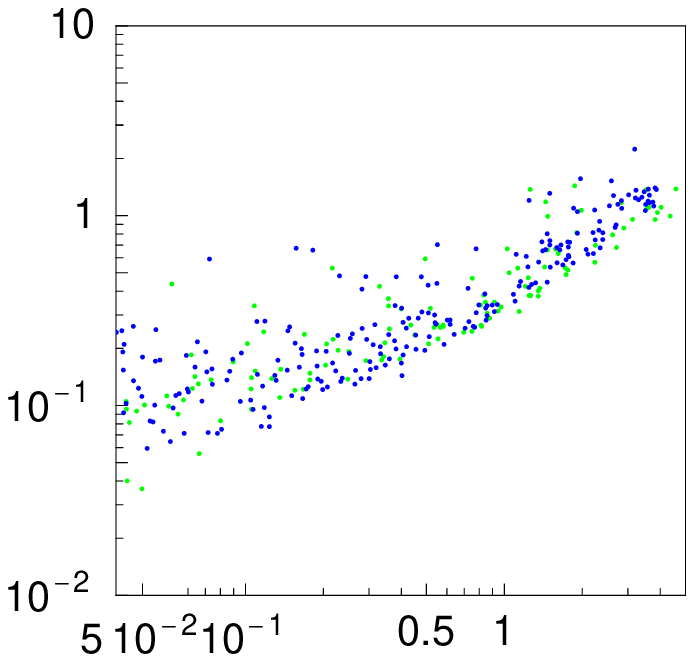,
                               height=7.3cm,width=7.cm}}}
\put(85,75){\makebox(0,0)[bl]{{\small
             b) Br($\mu^- \mu^+ \sum_i \nu_i$)/Br($\tau q q'$)}}}
\put(158,-3){\makebox(0,0)[br]{{$\tan^2(\theta_{atm})$}}}
\end{picture}
\caption[]{Testing the atmospheric angle.
           In case of the dark (bright) points  $\mu<(>)0$. 
           In the second figure we have taken 
           only those points with $|\sin 2 \theta_{\tilde b}| > 0.1$.}
\label{fig:TestAngleAtmB}
\end{figure}

As can be seen from the discussion in \sect{sec:approx} the
approximate formulas depend on the SUSY parameters, in particular on
the parameters of the MSSM chargino/neutralino sector. However, one
can see that the ratios of neutralino partial decay widths or of its
branching ratios is rather insensitive to the MSSM parameters. As has
been pointed out in \cite{nulong} the atmospheric angle depends on the
ratio of $\Lambda_\mu / \Lambda_\tau$. This ratio (at tree level) can
be obtained by taking the ratio $O^{cnw}_{L21} / O^{cnw}_{L31}$. This
leads immediately to the idea that the semi-leptonic branching ratios
into $\mu^\pm q \bar{q}'$ and $\tau^\pm q \bar{q}'$ should be related
to the atmospheric angle.  This is clearly demonstrated in
\fig{fig:TestAngleAtm} where we show the ratio of the corresponding
branching ratios as a function of $\tan^2(\theta_{atm})$. One sees
that present data imply that this ratio should be $O(1)$.
In particular, the relative yield of muons and taus will specify
whether or not the solution to the atmospheric neutrino anomaly occurs
for parameter choices in the ``normal'' range or in the ``dark-side'',
i.e. $\tan^2(\theta_{atm}) <1$ or $\tan^2(\theta_{atm})
>1$~\cite{darkside}.

The observed width of the band simply expresses the residual SUSY
parameter dependence, which comes partly from the 1-loop calculated
mass matrix and partly from the different contributions to these
decays.  If for some reason $|\sin 2 \theta_{\tilde b}| > 0.1$ the
dependence on the parameters in the 1-loop calculation is considerably
reduced because the sbottom/bottom loop dominates.  This leads to a
stronger correlation as seen in \figz{fig:TestAngleAtm}{b}.  The fact
that for $\mu > 0$ the band is wider is a consequence of the
discussion in the previous paragraph.

In \fig{fig:TestAngleAtmA} and \fig{fig:TestAngleAtmB} we show two
additional ratios which exhibit also a correlation with $\tan^2
\theta_{atm}$: Br$(\chiz{1} \to e^\pm \mu^\mp \sum_i \nu_i$) /
Br$(\chiz{1} \to \tau^\pm q \bar{q}'$) and Br$(\chiz{1} \to \mu^-
\mu^+ \sum_i \nu_i$) / Br$(\chiz{1} \to \tau^\pm q \bar{q}'$).  The
(nearly) maximal mixing of atmospheric neutrinos implies that several
other ratios of branching ratios are also fixed to within one order of
magnitude,
see Table 1 and \fig{fig:TestAngleSolPred}. 
%
\begin{figure}
\setlength{\unitlength}{1mm}
\begin{picture}(150,80)
\put(0,1){\mbox{\epsfig{figure=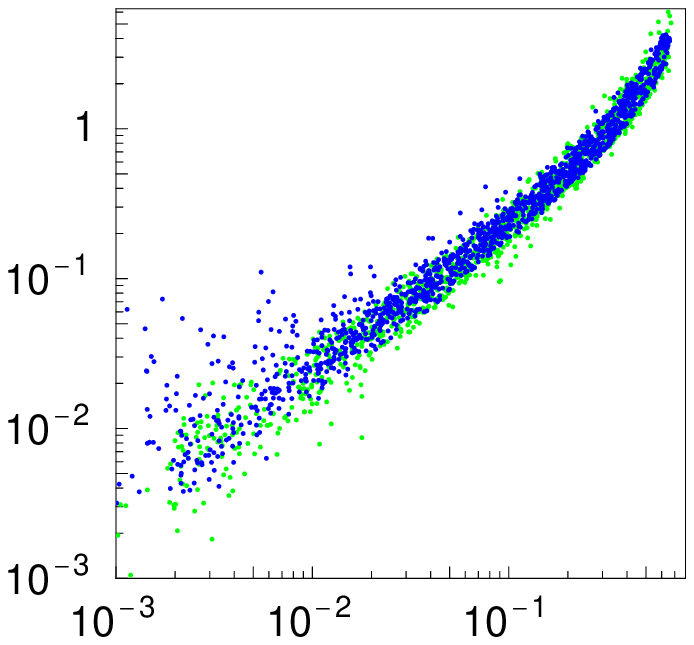,height=7.3cm,width=7.cm}}}
\put(0,76){\makebox(0,0)[bl]{{ a)}}}
\put(5,75){\makebox(0,0)[bl]{{\small Br($e q q'$)/Br($\tau q q'$)}}}
\put(70,-3){\makebox(0,0)[br]{{$U^2_{e3}$}}}
\put(88,1){\mbox{\epsfig{figure=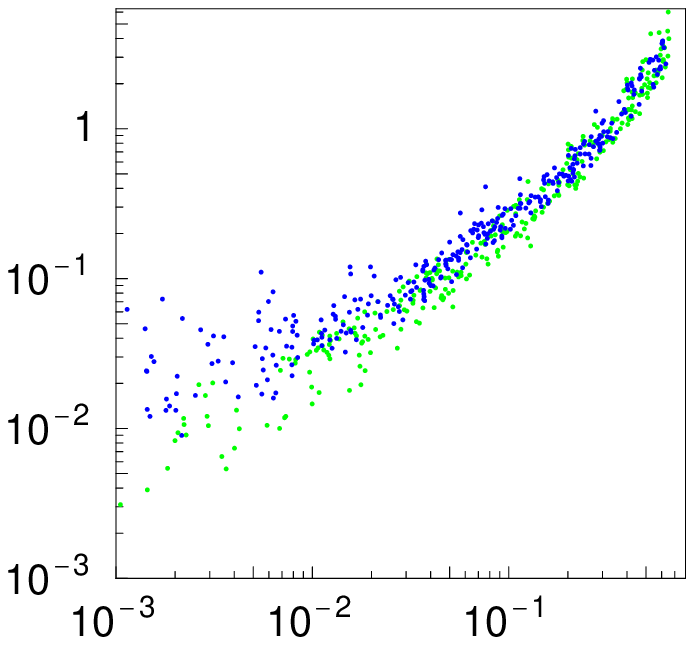,
                               height=7.3cm,width=7.cm}}}
\put(84,76){\makebox(0,0)[bl]{{ b)}}}
\put(89,75){\makebox(0,0)[bl]{{\small Br($e q q'$)/Br($\tau q q'$)}}}
\put(158,-3){\makebox(0,0)[br]{{$U^2_{e3}$}}}
\end{picture}
\caption[]{Testing the Chooz angle.
           In case of the dark (bright) points  $\mu<(>)0$. 
            In b) we have taken 
           only those points with $|\sin 2 \theta_{\tilde b}| > 0.1$.}
\label{fig:TestChooz}
\end{figure}

In this model the so-called Chooz-angle is given by $|\Lambda_e /
\Lambda_\tau|$~\cite{nulong} where we already have used the fact that
the atmospheric data implies $|\Lambda_\mu| \simeq |\Lambda_\tau|$.
The same discussion as in the previous paragraph is valid. This leads
automatically to the correlation between Br($\chiz{1} \to e^\pm q
q'$)/Br($\chiz{1} \to \tau^\pm q q'$) and $U^2_{e3}$ which is shown in
\fig{fig:TestChooz}. For $U^2_{e3} < 0.01$ the correlation is less
stringent because it implies that the tree level couplings have to be
rather small and therefore loop corrections are more important. Note
that existing reactor data~\cite{CHOOZ} give the constraint on
$U^2_{e3} \lsim 0.05$ at 90\% CL~\cite{latestglobalanalysis}. This in
turn implies an upper bound of $\sim0.2$ on this ratio of branching
ratios.

\begin{figure}
\setlength{\unitlength}{1mm}
\begin{center}
\begin{picture}(80,70)
\put(0,-5){\mbox{\epsfig{figure=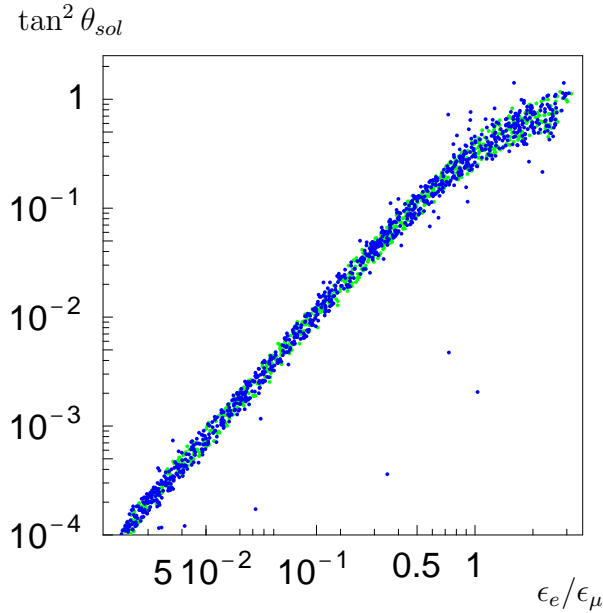,height=7.7cm,width=7.7cm}}}
\put(2,73){\mbox{$\tan^2 \theta_{sol}$}}
\put(71,-3){\mbox{$\epsilon_e/\epsilon_\mu$}}
\end{picture}
\end{center}
\caption{The solar mixing angle as a function of $\epsilon_e/\epsilon_\mu$.
         \label{fig:angles}}
\end{figure}

\begin{figure}
\setlength{\unitlength}{1mm}
\begin{picture}(150,80)
\put(0,1){\mbox{\epsfig{figure=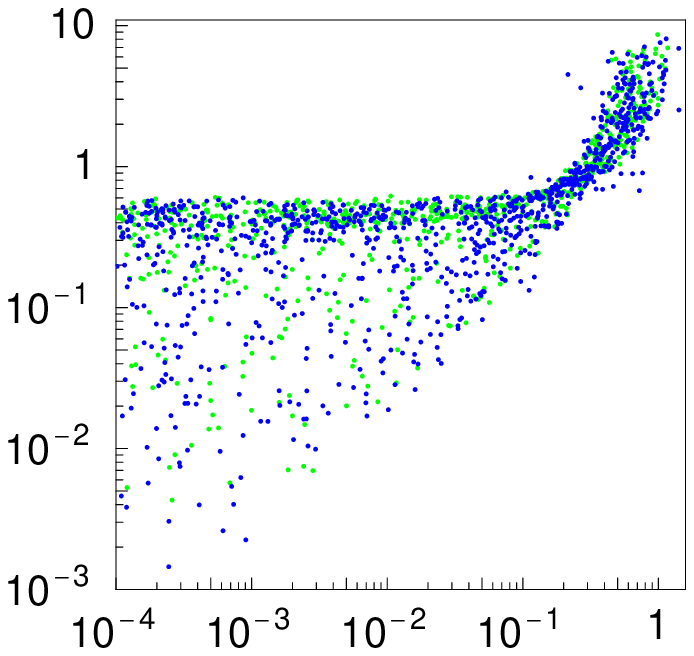,height=7.3cm,width=7.cm}}}
\put(0,76){\makebox(0,0)[bl]{{ a)}}}
\put(5,75){\makebox(0,0)[bl]{{\small Br($e \tau \nu_i$)/Br($\mu \tau \nu_i$)}}}
\put(70,-3){\makebox(0,0)[br]{{$\tan^2(\theta_{sol})$}}}
\put(88,1){\mbox{\epsfig{figure=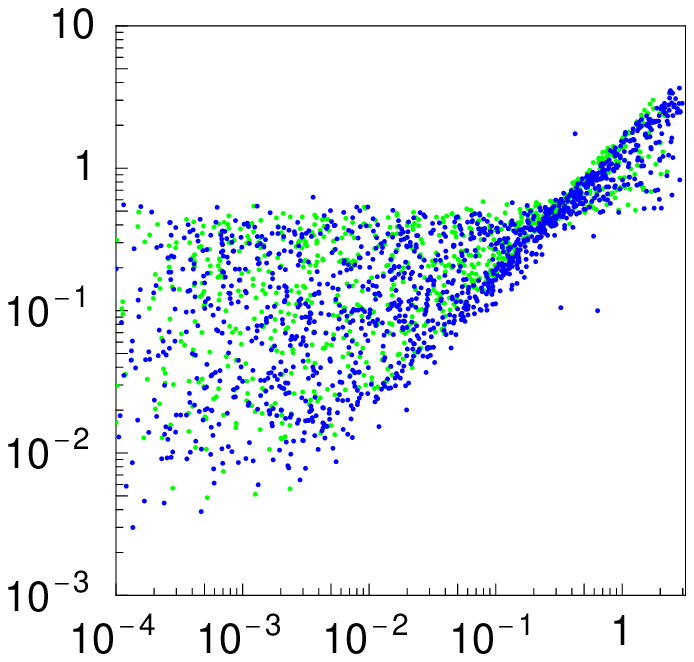,
                               height=7.3cm,width=7.cm}}}
\put(84,76){\makebox(0,0)[bl]{{ b)}}}
\put(89,75){\makebox(0,0)[bl]{{\small Br($e\tau \nu_i$)/Br($\mu \tau \nu_i$)}}}
\put(158,-3){\makebox(0,0)[br]{{$\tan^2(\theta_{sol})$}}}
\end{picture}
\caption[]{Testing the solar angle.
           In case of the dark (bright) points  $\mu<(>)0$. 
           In a) we have 
           taken $\epsilon_\mu \Lambda_\mu /(\epsilon_\tau \Lambda_\tau)$
     $> 0$. and in b) $\epsilon_\mu \Lambda_\mu /(\epsilon_\tau \Lambda_\tau)$
           $< 0$ }
\label{fig:TestAngleSol}
\end{figure}

\begin{figure}
\setlength{\unitlength}{1mm}
\begin{center}
\begin{picture}(150,80)
\put(0,-1){\mbox{\epsfig{figure=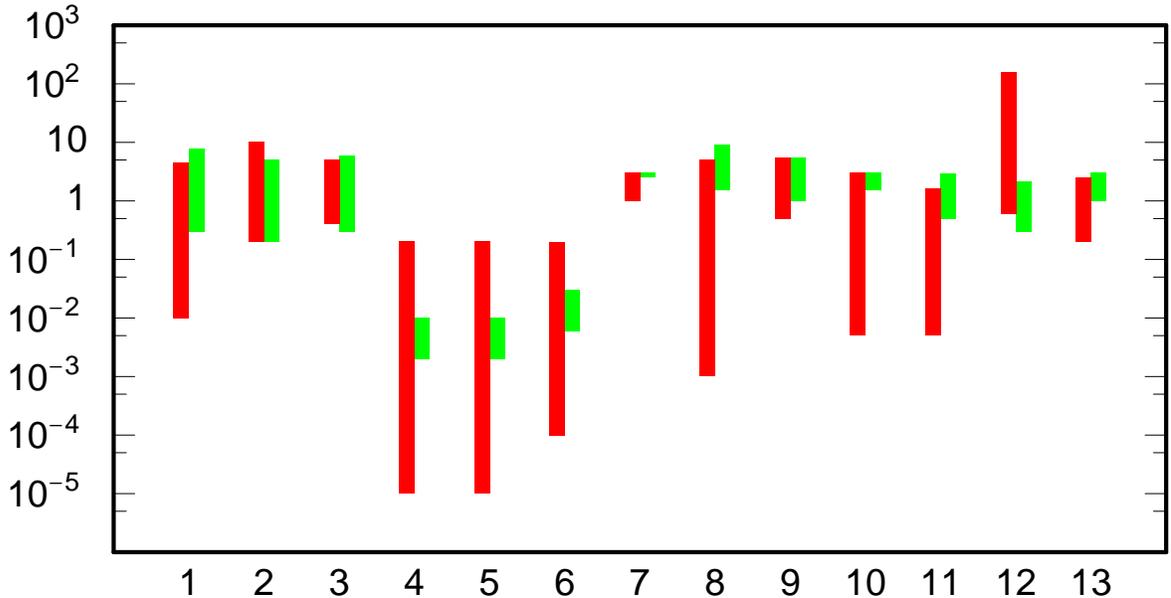,height=8cm,width=16.cm}}}
\end{picture}
\end{center}
\caption[]{Predicted ranges for the ratios of various branching ratios.
  The dark stripes are the ranges if one of the large mixing solutions
  (LMA, LOW or just-so) is realized in nature, the bright stripes are
  if SMA is realized in nature. The various ratios are given in the text. }
\label{fig:TestAngleSolPred}
\end{figure}

The discussion of the solar angle is more involved. As illustrated in
\fig{fig:angles} this angle is strongly correlated with
$\epsilon_e/\epsilon_\mu$ ratio.
In order to get information on the $\epsilon_i$ from neutralino decays
one must take into account that, as already mentioned, the solar angle
acquires a meaning only once the complete 1-loop corrections to the
mass matrix have been included. For an easier understanding we
focus on leptonic decays of the type
$\chiz{1} \to l^+_i l^-_j \nu_k$ with $i\ne j$ which depend on the
$\chiz{1}$-$W$-$l_{j,i}$ and $W$-$l_{i,j}$-$\nu_k$ couplings.
The way a correlation appears is non-trivial. To understand it note
that the couplings $W$-$l_i$-$\nu_j$ depend on the neutrino mixing,
since one must use the {\sl mass eigenstates} for the calculation of
the partial decay widths and not the electroweak eigenstates. In
addition the $\epsilon_i$ enter via the $\nu_j$-$S_k^\pm$-$l_i$ and
$\chiz{1}$-$S_k^\pm$-$l_i$ couplings.
Remarkably, despite the non-trivial way the $\epsilon_i$ parameters
enter here, one still has some residual correlation with $\epsilon_i$
ratios, as displayed in  \fig{fig:TestAngleSol}.
This figure shows that, although one does not get a strong correlation
in this case, one can still derive lower and upper bounds depending on
$\tan^2(\theta_{sol})$. For the favored
case~\cite{latestglobalanalysis} of the large mixing angle solution
one finds that Br($\chiz{1} \to e \tau \nu_i$)/Br($\chiz{1} \to \mu \tau
\nu_i$) is determined to be one to within an order of magnitude.
For the general bilinear \rp model the spread in
\fig{fig:TestAngleSol} is due to the lack of knowledge of SUSY
parameters. As will be shown in the next subsection, a much stronger
correlation appears once the SUSY parameters get determined.

In \tab{tab:TestSolar} we list upper and lower bounds on several
ratios of branching ratios which are required by the consistency of
the model. The values in the table are hardly dependent on the
solution for the solar neutrino problem.
\begin{table}
\label{tab:TestSolar}
\caption[]{Ratio of branching ratios as required by the consistency of
the model. In Br($q \bar{q} \sum_i \nu_i$) we have summed over $u$, $d$, 
and $s$.
Also in case of $\nu_i$ we have summed over all neutrinos.\\[-0.2cm]}
\begin{center}
\begin{tabular}{|l|c|c|} \hline
  Ratio & lower bound & upper bound  \\ \hline 
Br($ q \bar{q} \nu_i$) / Br($ c \bar{c} \nu_i$) &
                                                     2.5 & 6 \\
Br($ q \bar{q} \nu_i$) / Br($\mu^\pm q \bar{q}'$) &
                                                     0.1 & 3.5 \\
Br($ q \bar{q} \nu_i$) / Br($\tau^\pm q \bar{q}'$) &
                                                     0.1 & 3.5 \\
Br($ q \bar{q} \nu_i$) / Br($e^+ e^- \nu_i$) &
                                                       5 & 35 \\
Br($ q \bar{q} \nu_i$) / Br($e^\pm \mu^\mp \nu_i$) &
                                                    0.3 & 9.5 \\
Br($ q \bar{q} \nu_i$)/ Br($\mu^+ \mu^-  \nu_i$) & 
                                                     0.3 & 9 \\
Br($\mu^\pm q \bar{q}'$) / Br($\tau^\pm q \bar{q}'$) &
                             0.5 & 3 \\
Br($\mu^\pm q \bar{q}'$) / Br($\mu^+ \mu^- \nu_i$) &
                             1 & 5  \\
Br($\tau^\pm q \bar{q}'$) / Br($\mu^+ \mu^- \nu_i$) &
                             0.5 & 6.5  \\
Br($e^\pm \mu^\mp \nu_i$) / Br($\mu^+ \mu^- \nu_i$) &
                             0.4 & 1.6  \\
\hline
\end{tabular}
\end{center}
\end{table}

The values in \tab{tab:TestSolar} can be viewed as important consistency checks
of our model. However, one can also have observables which are able to
discriminate between large and small angle solution of the solar
neutrino problem.
In \fig{fig:TestAngleSolPred} we show how several ratios of neutralino
decay branching ratios can be used to discriminate between large and
small angle solution of the solar neutrino problem.
The numbers in  \fig{fig:TestAngleSolPred} correspond to the 
following branching ratios: 
1 \dots Br($ q \bar{q} \nu_i$) / Br($e^\pm \tau^\mp \nu_i$),
2 \dots Br($b \bar{b} \nu_i$) / Br($\mu^\pm \tau^\mp \nu_i$),
3 \dots Br($b \bar{b} \nu_i$) / Br($\tau^- \tau^+ \nu_i$),
4 \dots Br($e^\pm q \bar{q}'$) / Br($\mu^\pm q \bar{q}'$),
5 \dots Br($e^\pm q \bar{q}'$) / Br($\tau^\pm q \bar{q}'$),
6 \dots Br($e^\pm q \bar{q}'$) / Br($e^\pm \mu^\mp \nu_i$),
7 \dots Br($\mu^\pm q \bar{q}'$) / Br($e^\pm \mu^\mp \nu_i$),
8 \dots Br($\mu^\pm q \bar{q}'$) / Br($e^\pm \tau^\mp \nu_i$),
9 \dots Br($\tau^\pm q \bar{q}'$) / Br($e^\pm \mu^\mp \nu_i$),
10 \dots Br($\tau^\pm q \bar{q}'$) / Br($e^\pm \tau^\mp \nu_i$),
11 \dots Br($e^\pm \mu^\mp \nu_i$) / Br($e^\pm \tau^\mp \nu_i$),
12 \dots Br($e^\pm \tau^\mp \nu_i$) / Br($\mu^+ \mu^- \nu_i$), and 
13 \dots Br($\mu^\pm \tau^\mp \nu_i$) / Br($\tau^+ \tau^- \nu_i$).
In Br($q \bar{q} \sum_i \nu_i$) we have summed over $u$, $d$, and $s$.
Also for the case of $\nu_i$ we have summed over all neutrinos.

\subsection{After the  SUSY spectrum is measured}
 
In the previous section we have discussed the predictions which can be
established between neutralino decay branching ratios and neutrino
mixing angles before the first SUSY particle is discovered. 
Let us assume now that the entire spectrum has been measured with some
precision, e.g. at a future Linear Collider \cite{tesla,Martyn:1999tc}. 
As a typical example we discuss the point $M_2=120$~GeV,
$\mu=500$~GeV, $\tan \beta=5$, setting all scalar mass parameters to
$500$~GeV, and also the A-parameter is assumed to be equal for all
sfermions $A = -500$~GeV.  Note, that we have taken $\mu$ positive to
be conservative, as this corresponds to a 'worst-case' scenario."
There are at least two parameters which need to be measured precisely:
$\tan \beta$ and $|\sin 2 \theta_{\tilde b}|$ because the 1-loop mass
matrix is dominated by the sbottom/bottom loop if at least one of
these parameters is large.

\begin{figure}
\setlength{\unitlength}{1mm}
\begin{picture}(150,75)
\put(0,-1){\mbox{\epsfig{
          figure=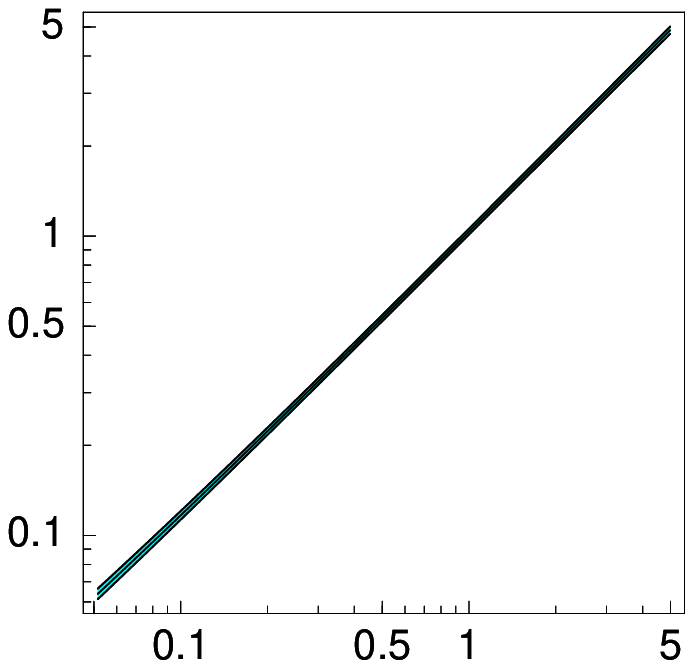,height=6.8cm,width=5.cm}}}
\put(0,68){\makebox(0,0)[bl]{{\small
                  a)   Br($\mu^\pm q \bar{q}'$)/Br($\tau^\pm q \bar{q}'$)}}}
\put(50,-3){\makebox(0,0)[br]{{$\tan^2 \theta_{atm}$}}}
\put(54,-2){\mbox{\epsfig{
        figure=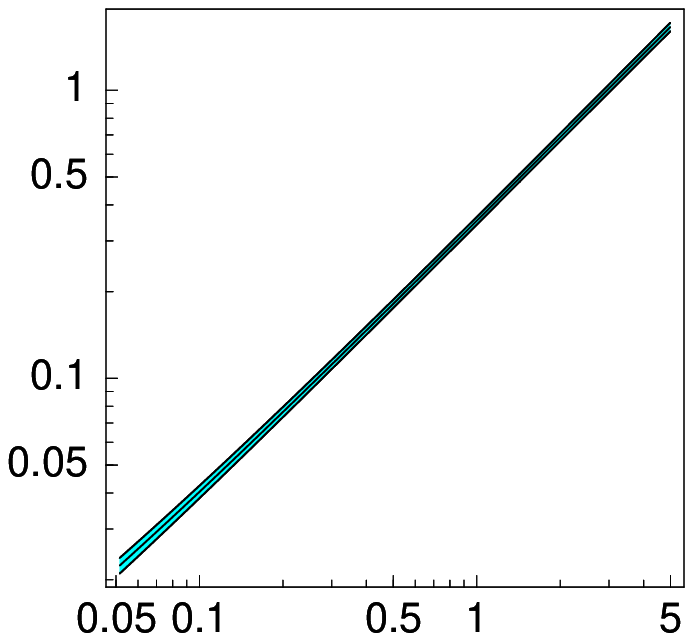,height=7.cm,width=5.cm}}}
\put(54,68){\makebox(0,0)[bl]{{\small
             b)   Br($e^\pm \mu^\mp \sum_i \nu_i$)/Br($\tau^\pm q \bar{q}'$)}}}
\put(104,-3){\makebox(0,0)[br]{{$\tan^2 \theta_{atm}$}}}
\put(108,-1){\mbox{\epsfig{
        figure=BrMuQQpTauQQpTanAtmsq.eps, height=6.8cm,width=5.cm}}}
\put(108,68){\makebox(0,0)[bl]{{\small
             c)   Br($\mu^+ \mu^- \sum_i \nu_i$)/Br($\tau^\pm q \bar{q}'$)}}}
\put(158,-3){\makebox(0,0)[br]{{$\tan^2 \theta_{atm}$}}}
\end{picture}
\caption[]{Correlations between $\tan^2 \theta_{atm}$ and ratios of
           branching ratio for the parameter point specified in the text
           assuming that $10^5$ neutralino decays have been measured.
           The bands correspond to an 1-$\sigma$ error.}
\label{fig:PostDictAtm}
\end{figure}

\begin{figure}
\setlength{\unitlength}{1mm}
\begin{center}
\begin{picture}(80,80)
\put(0,0){\mbox{\epsfig{figure=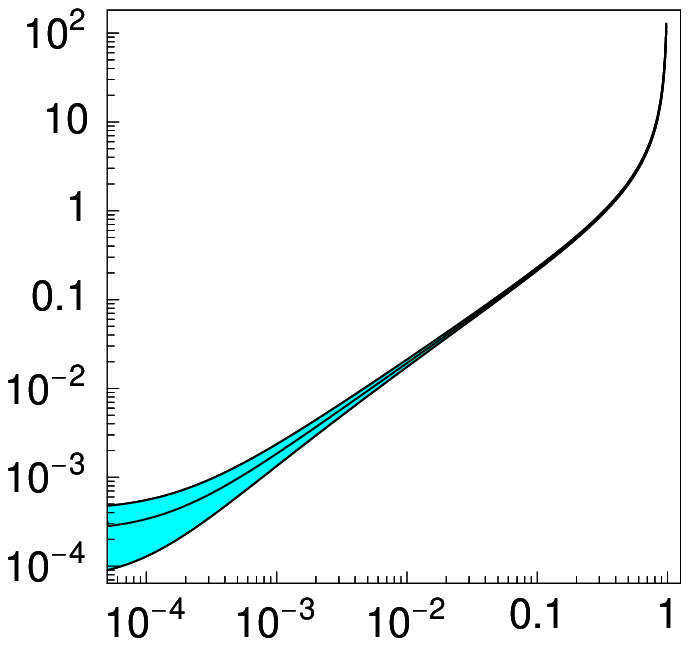,height=7.7cm,width=7.cm}}}
\put(2,76){\makebox(0,0)[bl]{{\small
                Br($e^\pm q \bar{q}'$)/Br($\tau^\pm q \bar{q}'$)}}}
\put(71,-3){\makebox(0,0)[br]{{$U^2_{e3}$}}}
\end{picture}
\end{center}
\caption[]{Correlation between $U^2_{e3}$ and the ratio
           Br($e^\pm q \bar{q}'$)/Br($\tau^\pm q \bar{q}'$) 
           for the parameter point specified in the text
           assuming that $10^5$ neutralino decays have been measured.
           The band corresponds to an 1-$\sigma$ error.}
\label{fig:ProsDictUe3}
\end{figure}

\begin{figure}
\setlength{\unitlength}{1mm}
\begin{center}
\begin{picture}(80,80)
\put(0,0){\mbox{\epsfig{figure=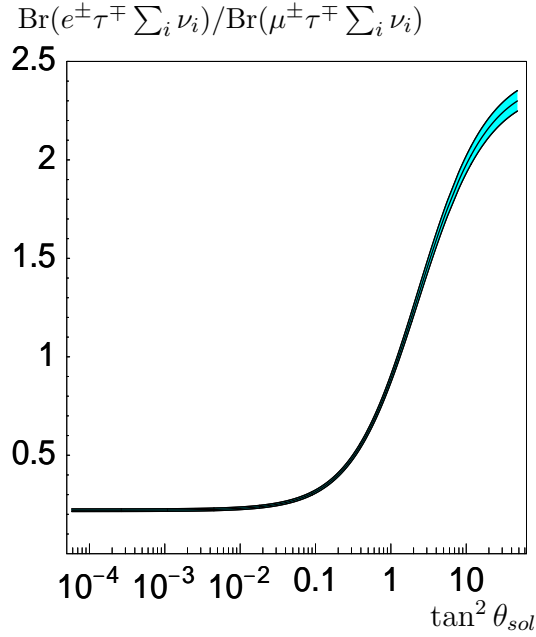,
                                     height=7.5cm,width=7.cm}}}
\put(2,76){\makebox(0,0)[bl]{{\small
     Br($e^\pm \tau^\mp \sum_i \nu_i$)/Br($\mu^\pm \tau^\mp \sum_i \nu_i$)}}}
\put(71,-3){\makebox(0,0)[br]{{$\tan^2 \theta_{sol}$}}}
\end{picture}
\end{center}
\caption[]{Correlation between $\tan^2 \theta_{sol}$ and the ratio
         Br($e^\pm \tau^\mp \sum_i \nu_i$)/Br($\mu^\pm \tau^\mp \sum_i \nu_i$)
           for the parameter point specified in the text
           assuming that $10^5$ neutralino decays have been measured.
           The band corresponds to an 1-$\sigma$ error.}
\label{fig:ProsDictSol}
\end{figure}

In \fig{fig:PostDictAtm} --- \fig{fig:ProsDictSol} the same
relationships as discussed above are displayed assuming that the
particle spectrum and the corresponding mixing angles are known to the
1\% level or better. In addition we have assumed that $10^5$
neutralino decays have been identified and measured.  Taking at the
moment only the statistical error this translates to a relative error
on the branching ratio Br$(X)$ of the form $1/\sqrt{10^5
  \mathrm{Br}(X)}$. It is clear from these figures that there exist
excellent correlations between the ratio of various branchings and
$\tan^2 \theta_{atm}$ as well as the parameter $U^2_{e3}$ probed in
reactor experiments. For the solar angle we observe a strong
dependence on $\tan^2 \theta_{sol}$ for the case of large mixing angle
solutions (LMA, LOW or vacuum) of the solar neutrino problem. 
For the small mixing angle MSW solution, even though the dependence on
$\tan^2 \theta_{sol}$ becomes, unfortunately, rather weak, the ratio
of branching ratios for Br($e^\pm \tau^\mp \sum_i \nu_i$)/Br($\mu^\pm
\tau^\mp \sum_i \nu_i$) is predicted with good accuracy for any
$\tan^2 \theta_{sol} \lsim 0.1$.

\section{Conclusions}
\label{sec:conclusion}

Supersymmetry with broken R Parity provides a predictive hierarchical
pattern for neutrino masses and mixings determined in terms of just
three independent parameters, assuming CP to be conserved in the
lepton sector.
This can solve the solar and atmospheric neutrino anomalies in a way
that allows leptonic mixing angles to be probed at high energy
colliders, providing an independent determination of neutrino mixing.
Taking into account data from atmospheric neutrino experiments we have
derived specific predictions for neutralino decays, illustrated in
\fig{fig:TestAngleAtm}, \fig{fig:TestAngleAtmA} and
\fig{fig:TestAngleAtmB}.
Probing the solar angle is more involved due to the intrinsic spread
in SUSY parameters, see \fig{fig:TestAngleSol}.
However, we have demonstrated that, with about $10^5$ neutralino
decays, and a determination of the spectrum of the theory to within
1\% or better, there are very stringent correlations between solar
neutrino-physics and neutralino-physics (see \fig{fig:ProsDictSol}).
We showed that several ratios of neutralino decay branching ratios can
be used to discriminate between large and small angle solutions of the
solar neutrino problem.  Therefore, the hypothesis that bilinear
R-parity violation is the origin of neutrino mass and mixing can be
easily ruled out or confirmed at future collider experiments. This
statement is actually more general, to the extent that the bilinear
model is an effective theory of a model where R-parity is violated
spontaneously.

\section*{Acknowledgments}

This work was supported by DGICYT under grants PB98-0693 and by the
TMR network grants ERBFMRXCT960090 and HPRN-CT-2000-00148 of the
European Union and by the European Science Foundation network grant N
86.  M.H.~was supported by the Marie-Curie program under grant No
ERBFMBICT983000 and W.P.~by a fellowship from the Spanish Ministry of
Culture under the contract SB97-BU0475382.

\newpage
\begin{appendix}
\section{Approximate Diagonalization of scalar mass matrix}
\label{app:approximate}

Let us assume that the following simplifying conditions hold
\begin{eqnarray}
&& |\epsilon_i \epsilon_j| \ll |B_k \epsilon_k|  \\
&& |v_i| \ll v_D, v_U \\
&& |\epsilon_i| \ll |\mu| \\
&& 2 |\cosa \left( B_i \epsilon_i - g_Z v_U v_i \right)
 + \sina \left( g_Z v_D v_i - \mu \epsilon_i \right)|\ll \no
&& 
| \cosaq \Delta M_{rad} + g_Z \left(v_U \cosa - v_D \sina \right)^2 
  + B_0 \mu \left( \cot \beta \cosa - \tan \beta \sina \right)^2 
  - \frac{\epsilon_i}{v_i} \left( v_D \mu - B_i v_U \right) |  \no \\
&& 2 |- \sina \left( B_i \epsilon_i - g_Z v_U v_i \right)
 + \cosa \left( g_Z v_D v_i - \mu \epsilon_i \right) | \ll\no
&& 
 | \sinaq \Delta M_{rad} + g_Z \left(v_U \sina + v_D \cosa \right)^2 
  + B_0 \mu \left( \cot \beta \sina + \tan \beta \cosa \right)^2 
  - \frac{\epsilon_i}{v_i} \left( v_D \mu - B_i v_U \right) | \no 
\end{eqnarray}
where $g_Z = (g^2+g'^2)/4$, 
$\Delta M_{rad} = 3 g^2 m_t^4 / (16 \pi^2 m^2_W \sin^2 \beta \sin^2 \theta)$
with  $\sin^2 \theta = (v_U^2 + v_D^2)/(v_U^2 + v_D^2+v_1^2 + v_2^2+v_3^2)$
and $\alpha$ is the
mixing angle that diagonalizes the upper left $2\times2$ sub-matrix
which corresponds to the usual Higgs mass matrix in the MSSM limit.
Under these approximations the mixing matrix reads
\begin{eqnarray}
R^{S^0} = \left( \begin{array}{ccccc}
   c_2 c_4 c_6 \sina& c_2 c_4 c_6\cosa  & s_2 & s_4 & s_6 \\
   c_1 c_3 c_5\cosa & -  c_1 c_3 c_5\sina & s_1 & s_3 & s_5 \\
   - s_1 \cosa - s_2 \sina & s_1 \sina - s_2 \cosa & c_1 c_2 & 0 & 0 \\
   - s_3 \cosa - s_4 \sina & s_3 \sina - s_4 \cosa & 0 & c_3 c_4  & 0 \\
   - s_5 \cosa - s_6 \sina & s_5 \sina - s_6 \cosa & 0 & 0  & c_5 c_6
  \end{array} \right) 
\end{eqnarray}
with
\begin{eqnarray}
&& s_{2 i} = \no
&& \frac{\cosa \left( B_i \epsilon_i - g_Z v_U v_i \right)
      + \sina \left( g_Z v_D v_i - \mu \epsilon_i \right) }
     {\cosaq \Delta M_{rad} + g_Z \left(v_U \cosa - v_D \sina \right)^2 
      + B_0 \mu \left( \cot \beta \cosa - \tan \beta \sina \right)^2 
      - \frac{\epsilon_i}{v_i} \left( v_D \mu - B_i v_U \right) } \no \\
&& s_{2 i - 1} = \no
&& \frac{- \sina \left( B_i \epsilon_i - g_Z v_U v_i \right)
         + \cosa \left( g_Z v_D v_i - \mu \epsilon_i \right) }
        {\sinaq \Delta M_{rad} + g_Z \left(v_U \sina + v_D \cosa \right)^2 
        + B_0 \mu \left( \cot \beta \sina + \tan \beta \cosa \right)^2 
        - \frac{\epsilon_i}{v_i} \left( v_D \mu - B_i v_U \right) } \no
\end{eqnarray}
where $i=1,2,3$ and $c_i = 1 - s^2_i /2$.  For the case where
$\frac{\epsilon_i}{v_i} \left( v_D \mu - B_i v_U \right)$ are equal
for all $i$, the $\epsilon_i \epsilon_j$ part in the mixing between
the sneutrinos becomes important. Therefore
\begin{eqnarray}
 R^{S^0} &\Rightarrow& R^\epsilon R^{S^0} \\
 R^\epsilon &=& \left( \begin{array}{ccccc}
    1 & 0 & 0 & 0 & 0 \\
    0 & 1 & 0 & 0 & 0 \\
    0 & 0 & \frac{-\epsilon_\tau}{\sqrt{\epsilon^2_1 + \epsilon^2_3}}
          & 0 & \frac{\epsilon_e}{\sqrt{\epsilon^2_1 + \epsilon^2_3}} \\
    0 & 0 & 0 & \frac{-\epsilon_\mu}{\sqrt{\epsilon^2_1 + \epsilon^2_2}}
              & \frac{\epsilon_e}{\sqrt{\epsilon^2_1 + \epsilon^2_2}} \\
    0 & 0 & \frac{\epsilon_e}{|\epsilon|}  & \frac{\epsilon_\mu}{|\epsilon|}
           & \frac{\epsilon_\tau}{|\epsilon|}
    \end{array} \right)
\end{eqnarray}
On the other hand if
$\frac{\epsilon_e}{v_1} \left( v_D \mu - B_1 v_U \right) =$
$\frac{\epsilon_\mu}{v_2} \left( v_D \mu - B_2 v_U \right) \ne$
$\frac{\epsilon_\tau}{v_3} \left( v_D \mu - B_3 v_U \right)$
and $|\epsilon_i \epsilon_j| \ll $
$| \frac{\epsilon_\mu}{v_2} \left( v_D \mu - B_2 v_U \right) -
 \frac{\epsilon_\tau}{v_3} \left( v_D \mu - B_3 v_U \right)|$
one has
\begin{eqnarray}
 R^\epsilon &=& \left( \begin{array}{ccccc}
    1 & 0 & 0 & 0 & 0 \\
    0 & 1 & 0 & 0 & 0 \\
    0 & 0 & \frac{-\epsilon_\mu}{\sqrt{\epsilon^2_1 + \epsilon^2_2}}
          & 0 & \frac{\epsilon_e}{\sqrt{\epsilon^2_1 + \epsilon^2_2}} \\
    0 & 0 & 0 & \frac{-\epsilon_e}{\sqrt{\epsilon^2_1 + \epsilon^2_2}}
              & \frac{-\epsilon_\mu}{\sqrt{\epsilon^2_1 + \epsilon^2_2}} \\
    0 & 0 & 0&0& 1
    \end{array} \right)
  \left( \begin{array}{ccccc}
    1 & 0 & 0 & 0 & 0 \\
    0 & 1 & 0 & 0 & 0 \\
    0 &0& c_7 & 0 & s_7 \\
    0 & 0 & 0 & c_8 & s_8 \\
    0 & 0 & -s_7 & - s_8 & c_7 c_8
    \end{array} \right)  \\
 s_7 &=& \frac{\epsilon_e \epsilon_\tau}
              { \frac{\epsilon_e}{v_1} \left( v_D \mu - B_1 v_U \right) -
                \frac{\epsilon_\tau}{v_3} \left( v_D \mu - B_3 v_U \right) } \\
 s_8 &=& \frac{\epsilon_\mu \epsilon_\tau}
              { \frac{\epsilon_e}{v_1} \left( v_D \mu - B_1 v_U \right) -
                \frac{\epsilon_\tau}{v_3} \left( v_D \mu - B_3 v_U \right) } \\
 c_{7,8} &=& 1 - \frac{s_{7,8}^2}{2}
\end{eqnarray}

We have checked, that the eigenvalues obtained with these mixing matrices
agree with those obtained in \cite{Grossman:2000ex} in lowest order of 
the R-parity breaking parameters.

\newpage
\section{Couplings}
Most of the couplings necessary for the calculation of the neutralino
decays have already been given in \cite{nulong} where also the
definition of the mixing matrices are given. The remaining couplings
involve $S^\pm_k$-$u_i$-$d_i$ and $\tilde q_j$-$q'_k$-$l^\pm_i$. Neglecting
the generation mixing among sfermions the
corresponding Lagrangian is given by:
\begin{eqnarray}
 {\cal L} &=&
       {\tilde d}_j \overline{u}_k \left( C^{\tilde d}_{Lk l_i} P_L
              + C^{\tilde d}_{Rk l_i} P_R \right)  l^+_i
      + {\tilde u}_j \overline{d}_k \left( C^{\tilde u}_{Lk l_i} P_L
              + C^{\tilde u}_{Rk l_i} P_R \right) l^-_i 
  \nonumber \\ &&
  + S^-_k \bar{d_i} \left( a_{S^-_ki} P_L + b_{S^-_ki} P_R \right)u_i
  + \mathrm{h.c.}
\end{eqnarray}
with
\begin{eqnarray}
  a_{S^-_ki} &=& h_D^{ii} R^{S^\pm}_{k1} \\
  b_{S^-_ki} &=& (h_U^{ii})^* (R^{S^\pm}_{k2})^* \\
 C^{\tilde d}_{Lk l_i}  &=& h_U^{kk}  ( R^{\tilde d}_{j1} )^*  V_{i2}^* \\
 C^{\tilde d}_{Rk l_i}  &=& - g  ( R^{\tilde d}_{j1} )^*  U_{i1}
            + (h_D^{kk})^*  ( R^{\tilde d}_{j2} )^*  U_{i2} \\
 C^{\tilde u}_{Lk l_i} &=& h_D^{kk}  ( R^{\tilde u}_{j1} )^*  U_{i2}^* \\
 C^{\tilde u}_{Rk l_i} &=& - g  ( R^{\tilde u}_{j1} )^*  V_{i1}
            + (h_U^{kk})^* ( R^{\tilde u}_{j1} )^*  V_{i2}
\end{eqnarray}

\end{appendix}

\newpage

\end{document}